\documentclass[debug]{rmaa}

\usepackage{paralist}

\usepackage{psfrag,color}

\usepackage[latin1]{inputenc}

\title{What do the star formation histories of galaxies tell us about the Starburst-AGN connection?} 

\author{
  J. P. Torres-Papaqui,\altaffilmark{1} 
  R. Coziol,\altaffilmark{1}
  I. Plauchu-Frayn,\altaffilmark{2}
  H. Andernach,\altaffilmark{1}
  and R. A. Ortega-Minakata\altaffilmark{1}}

\altaffiltext{1}{Departamento de Astronom\'{\i}a, Universidad de Guanajuato, Apartado Postal 144, 36000, Guanajuato, Gto, Mexico.}

\altaffiltext{2}{Instituto de Astronom\'{\i}a, Universidad Nacional Aut\'onoma de M\'exico, Campus Ensenada, Apdo. Postal 877, Ensenada, B.C.N. Mexico.}

\shortauthor{Torres-Papaqui et al.}
\shorttitle{The Starburst-AGN connection through the SFH}

\fulladdresses{
\item R. A. Ortega-Minakata, H. Andernach, R. Coziol and J. P. Torres-Papaqui: Departamento de Astronom\'{\i}a, Universidad de Guanajuato, Apartado Postal 144, 36000, Guanajuato, Gto, Mexico (rene, heinz, rcoziol, \& papaqui@astro.ugto.mx).
\item Ilse Plauchu-Frayn: Instituto de Astronom\'{\i}a, Universidad Nacional Aut\'onoma de M\'exico, Campus Ensenada, Apdo. Postal 877, Ensenada, B.C. Mexico (ilse@astrosen.unam.mx).}

\listofauthors{J. P. Torres-Papaqui,  R. Coziol, I. Plauchu-Frayn, H. Andernach, \& R. A. Ortega-Minakata}
\indexauthor{Torres-Papaqui, J.P.}
\indexauthor{Coziol, R.}
\indexauthor{Plauchu-Frayn, I.}
\indexauthor{Andernach, H.}
\indexauthor{Ortega-Minakata, R.A.}

\abstract
{We have determined the normal star formation histories (SFHs) for narrow emission line galaxies classified as star forming galaxies (SFGs), transition type objects (TOs), Seyfert~2s (Sy2s) and LINERs. The SFH varied with the activity type, following the mass of the galaxies and the importance of their bulge: LINERs reside in massive early-type galaxies, Sy2s and TOs in intermediate mass galaxies with intermediate morphological types, and SFGs are hosted in lower mass late-type spirals. Also, the maximum in star formation rate in the past was found to increase with the virial mass within the aperture (VMA). This correlation suggests that the bulges and the supermassive black holes at the center of galaxies grow in parallel, in good agreement with the M$_{\rm BH}$-$\sigma_*$ relation. }

\resumen
{
Determinamos las historias de formaci\'on estelar (SFHs) normal de galaxias con l\'ineas de emisi\'on angostas clasificadas como galaxias con formaci\'on estelar (SFGs), objetos de transici\'on (TOs), Seyfert~2 (Sy2s) y LINERs.  La SFH var\'ia con el tipo de actividad, siguiendo la masa de las galaxias y la importancia del bulbo: los LINERS residen en las galaxias m\'as masivas de tipos morfol\'ogicos m\'as tempranos, las Sy2s y los TOs en galaxias de masas intermediares con tipos morfol\'ogicos tambi\'en intermediares, y las SFGs est\'am hospedadas en galaxias espirales tard\'ias de masas menores. Tambi\'en, se encontro que la tasa m\'axima de formaci\'on estelar en el pasado aumenta con la masa virial en la apertura (VMA). Esto sugiere que los bulbos y los agujeros negros s\'uper masivos en los centros de galaxias se desarrolla en paralelo, de acuerdo con la relaci\'on M$_{\rm BH}$-$\sigma_*$.
}

\addkeyword{Galaxies: Activity}
\addkeyword{Galaxies: Star Formation}
\addkeyword{Galaxies: General}

\begin{document}
\maketitle

\section{Introduction}
\label{sec:intro}

According to the standard paradigm for AGNs, all massive galaxies possess a supermassive black hole (SMBH) at their center \citep[][]{Kormendy95,Richstone98}.  Consistent with this interpretation, during the last ten years it was established that the mass of the SMBH (M$_{\rm BH}$) is well correlated with the mass of the bulge, or stellar velocity dispersion ($\sigma_*$) of its host galaxy, the so called M$_{\rm BH}$-$\sigma_*$ relation \citep{Magorrian98,Ferrarese00,Gebhardt00,Haring04,Peterson05,Gultekin09,Graham11}. This correlation suggests that the formation of the SMBHs and the formation of the bulges of galaxies are intimately connected. What is the exact nature of this connection, however, is still a matter of debate. 

There are presently two different views on the origin of the M$_{\rm BH}$-$\sigma_*$ relation. The most popular one today \citep[e.g.][and references therein]{Page12} proposes that the formation of the bulge is self-regulated by the activity of the SMBH through an intricate feedback mechanism. In terms of star formation, this feedback is usually assumed to have had a negative effect, implying that in an AGN star formation is quenched when the SMBH reaches its state of maximum activity \citep[e.g.][]{Silk98,Hopkins06a,Croton06,Cattaneo09,Page12}. However, and despite the evidence presented in the literature supporting the quenching hypothesis, it is still not clear if the AGN feedback mechanism applies to all types of AGNs, and should thus be recognized as the principal process responsible for the M$_{\rm BH}$-$\sigma_*$ relation \citep{Heckman09,Wild10,Greene11,Mor12,TP12a}. 

The alternative point of view suggests that SMBHs grow in parallel with the bulges during the formation of the galaxies, when they pass through a succession of starbursts, that are induced in the nuclei by dissipative processes related with different types of merger events \citep{Silk81,Djorgovski87,Norman88,McLeod94,Priddey01,Carilli01,Page01,Mouri02,Heckman04,Farrah04,Gurkan04,Wild07,Sansigre08,Parra10,Rafferty11,Goto11,Cen12}. In the literature, this point of view is known and discussed in terms of the Starburst-AGN connection.

\begin{figure}
\includegraphics[width=\columnwidth]{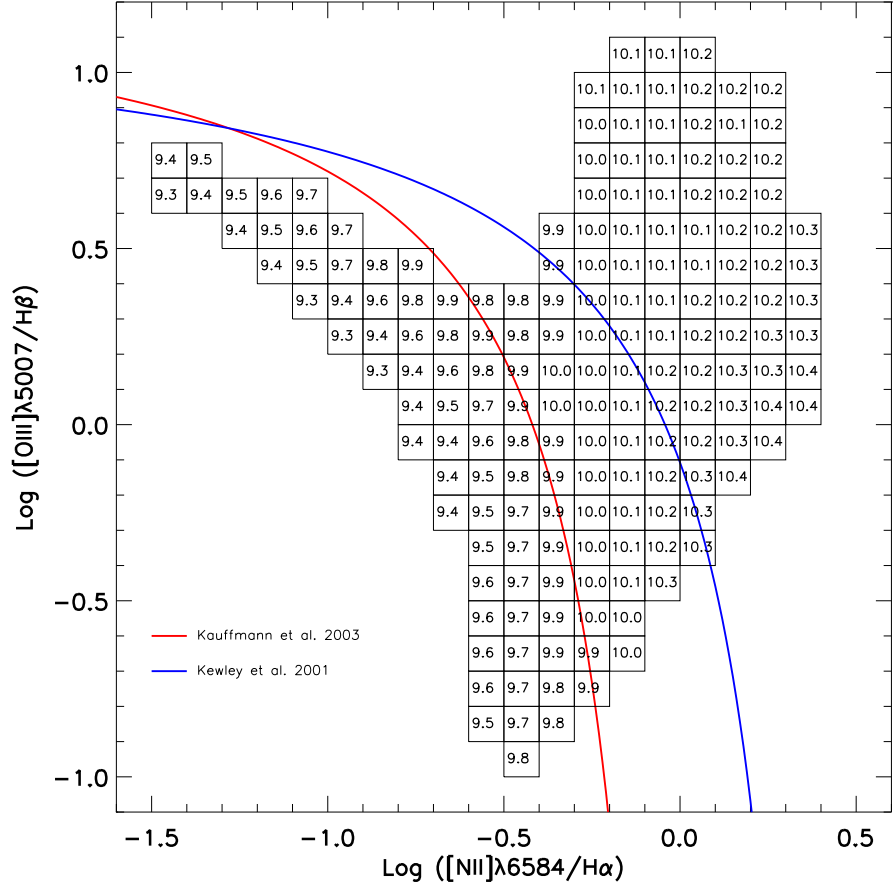}
\caption{BPT-VO diagram for the 229618 NELGs studied by \citet{TP12a}, showing the mean VMA as calculated in sub-samples of  equal size bins of line ratios. According to \citet{Kauffmann03} the red curve separates SFGs from AGNs, while the blue curve according to \citet{Kewley01} separates TOs from ``pure'' or dominating AGNs.}
\label{fig:01}
\end{figure}

In a previous article \citep{TP12a}, where we presented the results of a spectral analysis for a sample of 229618 Narrow Emission-Line Galaxies (NELGs) from the Sloan Digital Sky Survey, Data Release~7 \citep[SDSS, DR~7;][]{abazajian09}, we demonstrated that AGNs independently of their morphology or redshift always appear in galaxies with the highest VMA (the virial mass within the aperture of the SDSS fiber). This is illustrated in Figure~\ref{fig:01} where we show one standard diagnostic diagram \citep[hereafter identified as the BPT-VO diagram;][]{Baldwin81,Veilleux87} that separates star forming galaxies (SFGs) from AGNs.  The main separation criterion (red curve) was proposed by \citet{Kauffmann03}. In the BPT-VO diagram the NELGs trace a well known $\nu$-shaped distribution, with the SFGs on the left branch and the AGNs on the right branch. In Figure~\ref{fig:01} the whole sample was separated in equal size bins of 0.1 dex in [NII]$\lambda$5684/H$\alpha$ and [OIII]$\lambda$5007/H$\beta$, and the mean VMA within each bin was calculated. As can be observed, AGNs have a VMA of the order, or larger than $10^{10}$ M$_{\odot}$. The fact that the VMA is the highest in the AGNs suggests that when these galaxies formed they accumulated more mass at their center than any other type of galaxies (they have the highest gravitational potentials or gravitational binding energies). This relation is fully consistent with the existence of SMBHs at the center of all these galaxies \citep{TP12a}, in good agreement with the standard paradigm for AGNs.

On the other hand, in the same study we have also shown (by comparing the different power-laws traced by the continuum at 4800 \AA\ related with the H$\alpha$ emission-line fluxes) that more than 50\% of the AGNs in our sample seem to be buried in regions of intense star formation. In the literature these star forming AGNs are known as transition type objects (TOs), and can be found in Figure~\ref{fig:01} to the left of the empirical separation (blue curve) suggested by \citet{Kewley01}. The double nature of TOs--a galaxy with an AGN hidden behind circumnuclear star forming regions--seems fully consistent with what is observed in other more extreme cases of AGNs \citep[e.g.][]{Hamann92,Omont01,Isaak02,Meusinger02,Brinchmann04,Elbaz10}, which implies that evidence for intense star formation is common in AGNs \citep[][]{GonzalezDelgado01,Levenson01,Kauffmann03}. But is there a direct connection between the two phenomena? In \citet{TP12a} the TOs were found to have morphological types intermediate between the SFGs, located mostly in late-type spirals, and the LINERs, which are more numerous in early-type spirals and ellipticals. These variations in morphology seem to point toward a relation between the observed star formation activity and the growth of massive bulges. However, the role (if any) a SMBH may play in this process is unclear. For example, following the SMBH quenching hypothesis, the TOs could be AGNs in an early state of evolution (relative to the LINERS and Sy2s),  happening just before the AGN feedback effect. Alternatively, the TOs may yield evidence of coevolution, the SMBHs in these galaxies forming and developing at the same time as the bulges.

Finding evidence of truncation of star formation due to a SMBH in AGNs may be much more complicated than what is usually assumed. The problem is that AGNs are most frequently found in early-type galaxies, and it is well known that these galaxies form their stars in a way that rapidly deplete their gas reservoir, shortening their phase of active star formation \citep{Sandage86,Coz11}. These galaxies did not suffer from a truncation of star formation, but rather experienced higher astration rates during their formation, which also explains why they have so massive bulges. How then can we distinguish between the two phenomena, astration or truncation? Indeed, how do we define truncated star formation? As a working definition, we propose the following: a galaxy shows evidence for truncated star formation when, anytime during its formation, we observe a dramatic and unusual reduction of its star formation rate. Applying this working definition suggests one way how to search for evidence of a quenching effect due to a SMBH: the evidence for truncated star formation should be a step function in a plot of the star formation rate as a function of time. Consequently, this galaxy should show a star formation history (SFH) that is unusual when compared to the ``normal'' star formation histories of galaxies with the same morphology. 

Consistent with the above definition, we have recently developed a new method based on stellar population synthesis that allowed us to detect evidence of unusual SFHs in nearby galaxies forming compact groups \citep{Plauchu-Frayn12}. But although our observations seemed, at first, to be fully consistent with the truncated star formation phenomenon, confirming, in a way, the common impression about galaxies in compact groups, we have found that if one considers the transformation of these galaxies into earlier morphological types this phenomenon looks more like a sudden increase of astration rate, related with the recent formation of these dense structures. It seems essential, therefore, that before concluding in favor of a truncation of star formation in any AGN one must take into account how the hosts of these galaxies form normally, and, in particular, obtained their final morphology. 

Having in hand the 229618 NELGs from our previous analysis \citep{TP12a}, we decided to tackle the above problems by studying in detail their SFHs. We hope that by determining how galaxies with different nuclear activity types formed their stars, one can get new insights about the Starburst-AGN connection and the origin of the M$_{\rm BH}$-$\sigma_*$ relation. 

The organization of this article is the following. In Section~\ref{sec:sample} we present our sample, and explain the criteria applied to divide it into six activity groups. Having already determined the morphologies for all these galaxies in \citet{TP12a}, we show in Section~\ref{sec:sample} why this important characteristic of galaxies must be taken into account in our analysis of the SFHs. Other parameters introduced in Section~\ref{sec:sample} that are necessary for our analysis are the stellar velocity dispersions, used to estimate the VMAs, the stellar masses of the galaxies and the absolute magnitudes in B. In Section~\ref{sec:method} we explain how the SFHs were obtained and introduce our method of comparison. The results of our analysis can be found in Section~\ref{sec:results}, which are then discussed in Section~\ref{sec:discussion}. This is followed by a brief summary and conclusions in Section~\ref{sec:conclusion}.

\section{Physical characteristics of the sample}
\label{sec:sample}

\subsection{Description of the sample and data used in our analysis}

The sample of galaxies employed for our SFH analysis is the same as in our previous study \citep{TP12a}. It consists of 229618 NELGs from SDSS, DR~7, with redshifts $z \leq 0.25$, for which all the spectra were downloaded from the Virtual Observatory (VO) service\footnote{http://www.starlight.ufsc.br}. In the VO, the spectra are already processed through the spectral synthesis code \begin{scriptsize}STARLIGHT\end{scriptsize} \citep{CidFernandes05}, which produces a stellar template for each galaxy.  After this template is subtracted from the original spectrum a pure emission-line spectrum is obtained, where the fluxes of various lines can be measured with high precision. Our nuclear activity classification was done using three standard diagnostic diagrams: [OIII]$\lambda$5007/H$\beta$ vs. [NII]$\lambda$6584/H$\alpha$, [OIII]$\lambda$5007/H$\beta$ vs. [SII]$\lambda\lambda 6717,6731$/H$\alpha$ and [OIII]$\lambda$5007/H$\beta$ vs. [OI]$\lambda 6300$/H$\alpha$. We have used only galaxies with emission lines that have a signal-to-noise ratio $S/N \geq$ 3 and an adjacent continuum with $S/N \geq$ 10.

\begin{figure}
\includegraphics[width=\columnwidth]{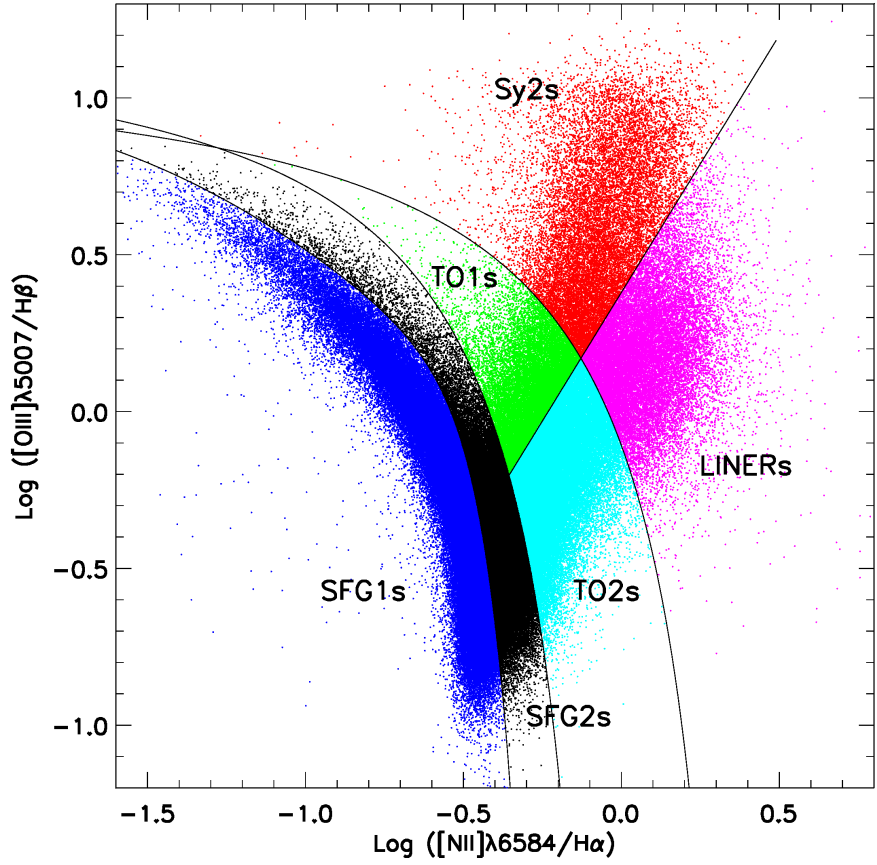}
\caption{BPT-VO diagram for the NELGs presenting our six different nuclear activity groups: SFG1s in blue, SFG2s in black, TO1s in green, TO2s in light blue, Sy2s in red and LINERs in magenta.}
\label{fig:02}
\end{figure}

For our study we have separated the NELGs in six activity groups as presented in Figure~\ref{fig:02}. The criterion used to differentiate SFGs from AGNs is that of \citet{Kauffmann03}. For the separation between AGNs and TOs we used the criterion proposed by \citet{Kewley01}. Our method to distinguish between LINERs and Sy2s is explained in \citet{TP12a}. By extending this distinction to the TOs, we produced two different groups, the TO1s, which are high excitation NELGs, like the Sy2s,  and the TO2s, which are low excitation NELGs, like the LINERs. Follow\-ing \citet{TP12b} we also distinguished between the SFG2s, that show an apparent excess of nitrogen emission and the SFG1s, which show no such excess. The number of galaxies in each group, N,  is shown in column~2 of Table~\ref{tab:01}. From this table we deduce that the dominant activity type in our sample of NELGs is SFGs (53.5\%), with slightly more SFG1s than SFG2s. The second most dominant type (28.8\%) are the TOs, with almost three times more TO2s than TO1s. NELGs where the AGN is predominant are consequently rare, forming only 17.7\% of our whole sample, with slightly more LINERs  (1.4 times more) than Sy2s. As we already emphasized in our introduction, this result implies that intense star formation is a very common phenomenon in AGNs at low redshifts \citep[see also][]{Lawrence85,Dultzin-Hacyan88,GonzalezDelgado01,Levenson01,Elbaz10}. 

\begin{table*}[!t]\centering
  \small
  \newcommand{\DS}{\hspace{6\tabcolsep}} 
  \begin{changemargin}{0.5cm}{0.5cm}
    \caption{Mean physical characteristics of the NELGs separated into six nuclear activity groups} \label{tab:01}
    \setlength{\tabnotewidth}{0.95\linewidth}
    \setlength{\tabcolsep}{1.0\tabcolsep} \tablecols{8}
    \begin{tabular}{lccccccc}
      \toprule
(1)&(2)&(3)&(4)*&(5)&(6)&(7)*&(8)\\
    Act.     & N       &  $\sigma_{ap}$   & VMA          & T   &  M$_B$ & M$_\star$     &  M$_\star$/L$_{\rm B}$\\
    group    &         &  (km s$^{-1}$)   & (M$_\odot$)  &     &        &  (M$_\odot$)  &  (M$_\odot$/L$_\odot$)    \\
    \hline
    SFG1     & 67513     &  92.1     &  9.59    &  4.1     & $-18.90$  & 10.14 & 0.91 \\
    SFG2     & 55238     & 110.3    &  9.84    &  3.5     & $-19.30$  & 10.60 & 1.43 \\
    TO1      & 18285     & 128.1    &  9.98    &  2.9     & $-19.13$  & 10.66 & 2.09 \\
    TO2      & 47949     & 134.9    & 10.05    &  2.7     & $-19.35$  & 10.81 & 2.23 \\
    Sy2      & 16970     & 136.3    & 10.09    &  2.8     & $-19.44$  & 10.81 & 2.44 \\
    LINER    & 23663     & 174.7    & 10.22    &  1.8     & $-19.26$  & 11.01 & 3.77 \\
      \bottomrule
    \end{tabular}
{\small In columns followed by an asterisk, values are given in logarithm.}
  \end{changemargin}
\end{table*}

A crucial characteristic of galaxies that was obtained from the templates produced by \begin{scriptsize}STARLIGHT\end{scriptsize} is the stellar velocity dispersion as measured within the SDSS fiber aperture, $\sigma_{ap}$. Note that \begin{scriptsize}STARLIGHT\end{scriptsize} does not take into account rotation. However, due to the small projected diameter of the SDSS fiber in the sky (3 arcseconds) and the distances of the galaxies in our sample ($z \leq 0.25$) the corresponding physical sizes in the galaxies always fall well within their bulges \citep{TP12a}, which suggests that $\sigma_{ap}$ is determined by the dynamics of the stars forming the bulge \citep[a pressure supported structure;][]{BBF92}. Within these conditions, and considering the resolution of the stellar library used by \begin{scriptsize}STARLIGHT\end{scriptsize}, $\sigma_{ap}$ rapidly converges to $\sigma_*$, with an uncertainty smaller than 14\%  \citep{Bernardi03}. Using $\sigma_{ap}$, we determine the VMA as the virial mass within the physical radius, $R$, which corresponds to the aperture of the SDSS fiber:

\begin{equation}
{\rm VMA} = \frac{R \sigma_{ap}^2}{G}
\end{equation}

The means for the velocity dispersions, $\sigma_{ap}$, and the VMA in solar mass (M$_\odot$) as calculated using Equation~1 are given in column 3 and 4, respectively, of Table~\ref{tab:01}. 

A critical phenomenon that must be taken into account during our analysis of the SFHs is the fact that the different activity types are correlated with the galaxy morphologies \citep{Lei00,Coz11,TP12a}. In \citet{TP12a}, the morphologies of all the galaxies were determined using the method developed by \citet{Shimasaku01} and \citet{Fukugita07}. This morphological classification uses the photometric colors as defined in the $ugriz$ photometric system of SDSS\footnote{http://casjobs.sdss.org}, and the inverse concentration index, which is the ratio of the Petrosian radii. After applying a K-correction \citep{Blanton07}, correlations were established with the standard Hubble morphological types \citep{deVauculeurs91}, using the morphological index T: (E,$-5$), (E/S0,\ $-2$), (S0,\ $0$), (S0/Sa, 0.5), (Sa, 1.0), (Sab, 2.0), (Sb, 3.0), (Sbc, 4.0), (Sc, 5.0), (Scd, 6.0), (Sd, 7.0), (Sd/Sm, 9.0), and (Im, 10). The variation of morphology with activity type can be appreciated in Table~\ref{tab:01}, column~5, where we present the average values for T as determined in each activity group.  

\begin{figure}
\includegraphics[width=\columnwidth]{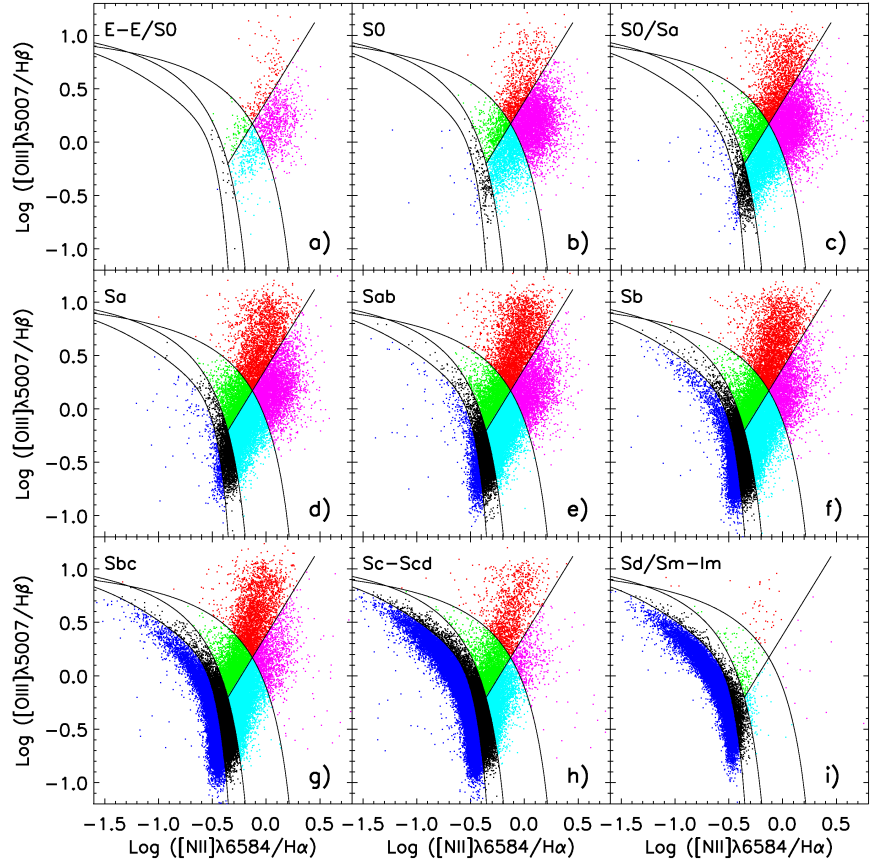}
\caption{BPT-VO diagrams for the NELGs separated by morphology groups. The percentage of the total population with different activity types in each morphological bin are presented in Table~\ref{tab:02}. The color code identifying the activity group is the same as in Figure~\ref{fig:02}.}
\label{fig:03}
\end{figure}

In Figure~\ref{fig:03} we show what happens in the BPT-VO diagram when we separate the activity groups in sub-samples with different morphological types. For each morphological bin presented in Figure~\ref{fig:03} the percentage of galaxies in the sub-samples are indicated in Table~\ref{tab:02} (columns 2 to 10). For the AGNs, the LINERs are more common in early-type galaxies, being more frequent in the S0-S0/Sa bins, while the Sy2s become more frequent in early spirals, reaching a maximum in the Sab-Sb bins. For the TOs we observe a trend toward later morphological types: the TO2s and TO1s are most frequent in intermediate-type spirals, reaching their maximum respectively in the Sb bin for the TO2s and Sbc bin for the TO1s. The SFGs continue the trend towards later-types, both the SFG2s and SFG1s reaching their maximum (mostly concentrated) in the Sc-Scd bin. 

\begin{table*}[!t]\centering
  \small
  \newcommand{\DS}{\hspace{6\tabcolsep}} 
  \begin{changemargin}{-1.3cm}{-1.3cm}
    \caption{Percentage of galaxies in each bin of morphology \\ as presented in Figure~3} \label{tab:02}
    \setlength{\tabnotewidth}{0.95\linewidth}
    \setlength{\tabcolsep}{1.2\tabcolsep} \tablecols{10}
    \begin{tabular}{lccccccccc}
      \toprule
(1)&(2)&(3)&(4)&(5)&(6)&(7)&(8)&(9)&(10)\\
Act.  & E-E/S0 &   S0    & S0/Sa   &   Sa    &  Sab   &   Sb     &   Sbc   & Sc-Scd   & Sd/Sm-Im \\
group & (\%) & (\%)& (\%)& (\%)& (\%)& (\%)& (\%)& (\%)& (\%)\\
      \midrule
LINER   & 3.59 & {\bf 20.16} & {\bf 22.68} & 19.0 & 14.85 & 10.95 &  6.51 &  2.16  &  0.07 \\
SY2     & 0.95 &  5.49 & 10.91 & 16.51 & {\bf 20.48} & {\bf 20.64} & 16.92 &  7.84  &  0.26 \\
TO2     & 0.70 &  4.15 &  8.77 & 14.25 & 19.42 & {\bf 21.73} & {\bf 21.48} &  9.28  &  0.22 \\
TO1     & 0.44 &  3.19 &  7.47 & 12.46 & 17.11 & {\bf 20.49} & {\bf 22.55} & 15.46  &  0.83 \\
SFG2    & 0.01 &  0.11 &  0.55 &  1.72 &  4.54 & 10.31 & {\bf 24.77} & {\bf 40.63}  & 12.23 \\
SFG1    & 0.00 &  0.14 &  0.61 &  1.86 &  4.81 & 11.03 & {\bf 27.54} & {\bf 44.25}  & 13.20 \\
      \bottomrule
\end{tabular}
{\small The fractions in bold identify the two most frequent morphologies.}
\end{changemargin}
\end{table*}

The next parameter listed in Table~\ref{tab:01}  (column~6) is the absolute magnitude in B, which was obtained using the Johnson-B band magnitudes synthesized from the SDSS magnitudes \citep{fukugita96}. Considering the low limit in redshift of our study, no special cosmology was applied to calculate this value, adopting for the Hubble constant H$_0 = 75$ km s$^{-1}$ Mpc$^{-1}$. 

To determine the mass in stars of all the galaxies in our sample, we used the empirical relation established by \citet{Bell03} between the mass-luminosity ratio, M$_\star$/L$_B$, and the $B-R$ colors. For the color we used the transformation of \citet{fukugita96}:  $(B-R) = 1.506(g-r)+ 0.370$, where $(g-r)$ is the color in the $ugriz$ photometric system of SDSS. This relation has a systematic error of $\sim 0.05$ mag \citep{Bell03,fukugita96}. The total mass in stars, M$_\star$, of all the galaxies in our sample were then deduced by multiplying the mass-luminosity ratio by the luminosity  \citep{Kauffmann03b}. The average values, in solar mass, of the NELGs in our sample are presented in column~7 of Table~\ref{tab:01}. In column~8 we have included the mean mass-luminosity ratios (M$_\star$/L$_B$). Note that the mean mass-luminosity ratios of some of the early-type galaxies in our sample seem somewhat low \citep[e.g.][]{Faber79,Roberts94}. This could suggest that the galaxies in our sample are slightly bluer than normal for their morphology. This feature is consistent with our selection criterion, that favors galaxies with strong emission lines.

\subsection{Statistical comparisons of the physical characteristics}

\begin{figure}
\includegraphics[width=\columnwidth]{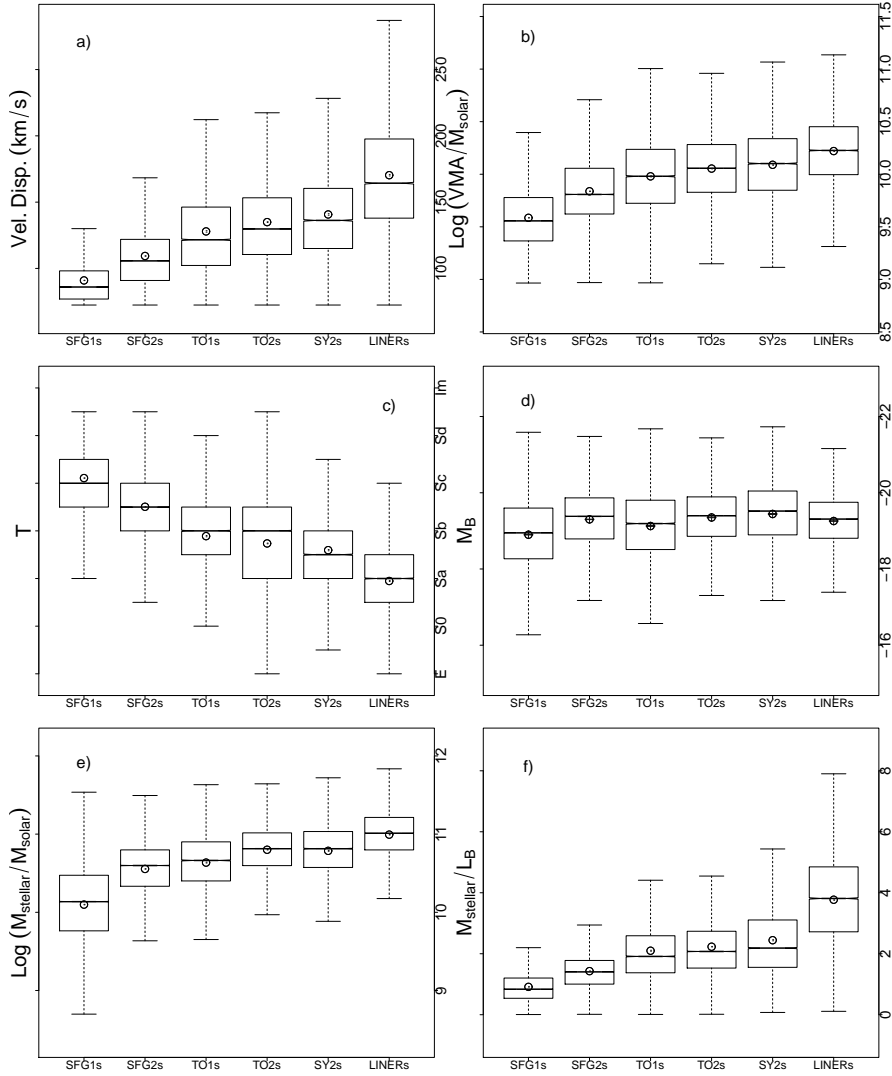}
\caption{Box-whisker plots comparing the characteristics of galaxies with different activity types: a) velocity dispersion, b)VMA, c) morphology, d) absolute magnitude in B, e) total mass of stars, f) mass-luminosity ratio. The lower side of the box is the 25th percentile, $Q_1$, and the upper side is the 75th percentile, $Q_3$. The whiskers correspond to $Q_1 - 1.5\times IQR$ and $Q_3+ 1.5\times IQR$. The median is shown as a bar and the mean as a circle. Although barely visible, the box-whisker plots include notches around the medians (see explanations in the text).}
\label{fig:04}
\end{figure}

In Figure~\ref{fig:04} we compare the physical characteristics of the galaxies with different activity types. Already from the box-whisker plots, the statistical significance of the differences found between the activity groups can be appreciated from the notches. The notches are V-shape regions, drawn around the medians, that have a width proportional to the  interquartile range, $IQR$ (the difference between the upper quartile and lower quartile),  and inversely proportional to the square root of the size of the sample, $N$ (that is, the width of one notch is given by $\pm 1.58 \times IQR/\sqrt{N}$). Comparing two samples, consequently, no overlapping notches implies that the medians of the two samples have a high probability of being different. Note that when a sample has a small dispersion (small $IQR$) and a high number of galaxies the width of the notches become very small, in fact, barely visible for some samples in Figure~\ref{fig:04}. This makes the test based on the comparison of notches even more conclusive. 

\begin{figure}
\includegraphics[width=\columnwidth]{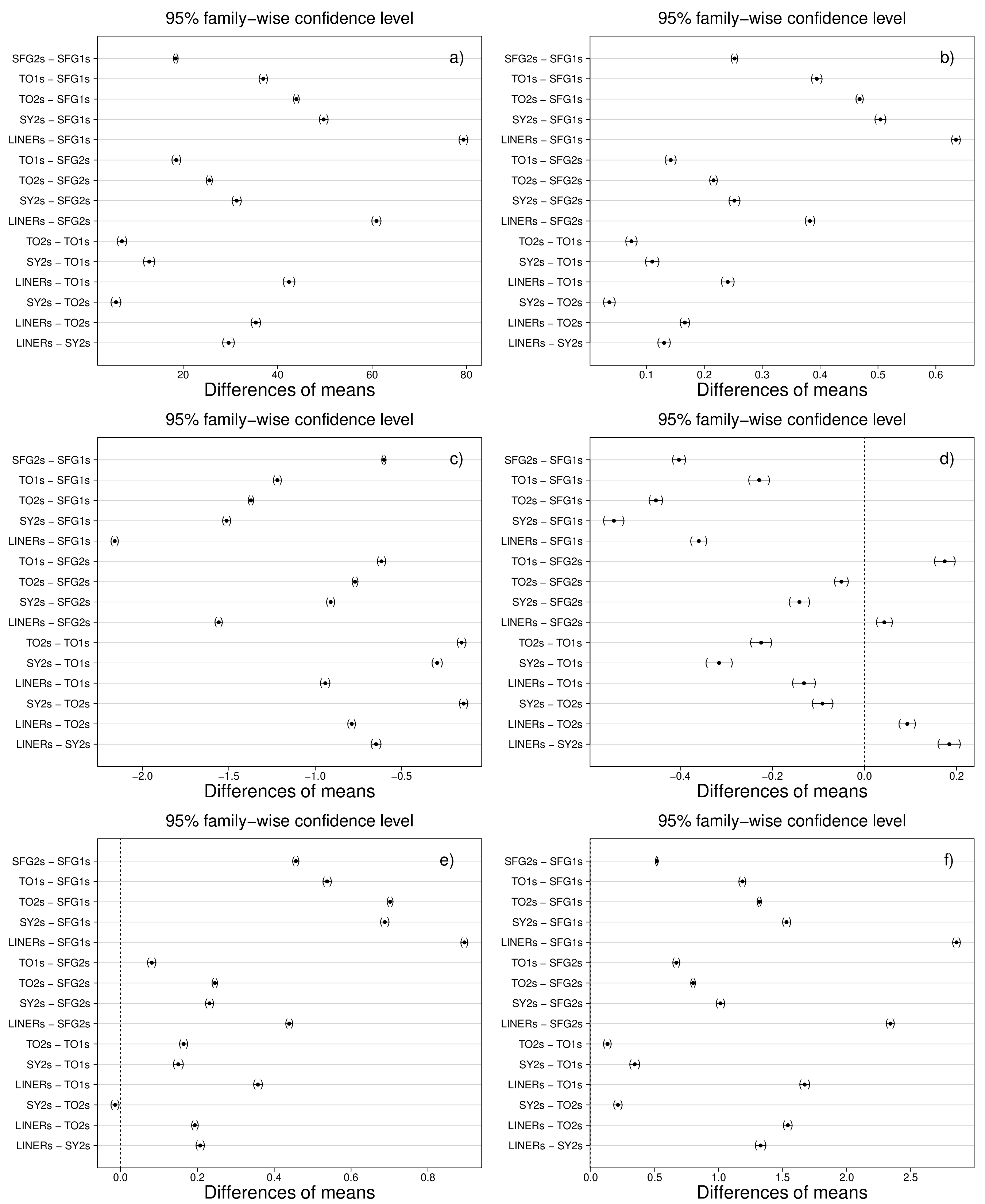}
\caption{Confidence intervals to estimate the statistical significance of the differences of the means: a) velocity dispersion, b) VMA, c) morphology, d) absolute magnitude in B, e) total mass of stars, f) mass-luminosity ratio. Confidence intervals including zero indicate no statistically significant differences at a 95\% level of confidence. The farther from zero, the more significant the difference in means. The smallness of the confidence intervals is due to the small dispersions ($IQR$) and large sizes of the samples. This makes the results of the statistical tests extremely significant.}
\label{fig:05}
\end{figure}

To confirm the statistical significance of the differences observed between the samples in Figure~\ref{fig:04}, we have traced in Figure~\ref{fig:05} the confidence intervals for the pairwise comparisons of the subsample means (that is, comparing simultaneously every mean with every other mean). This test assumes the null hypothesis takes a linear form, as in Tukey's method \citep[see explanations in][]{Hothorn08}. A confidence interval is obtained by calculating the difference between the means (for example, $M_1 - M_2$) then adding (for the upper limit) or subtracting (for the lower limit) the standard error (usually determined assuming all the samples have the same variance), which is multiplied by the value of a t-distribution at the level of confidence considered (usually 95\%). Consequently, there is a 95\% chance that the difference between the two means is located within the confidence interval calculated. Therefore, if $M_1 - M_2 > 0$, and the confidence interval does not include 0, one can securely state that the mean $M_1$ is higher than the mean $M_2$ (with a 5\% chance of being wrong). Note that in our analysis we have used a new test, the Max-t test  \citep{Hothorn08,Herberich10} that does not assume the variances are the same (heteroscedasticity) or that the sizes of the groups are comparable. The application of this test is relatively easy using the R software\footnote{http://CRAN.R-project.org}, and we have found it to be quite efficient for very large samples as considered in the present analysis. 

The box-whisker plots in Figure~\ref{fig:04}a show that the velocity dispersion of the galaxies significantly increases from the SFGs to the LINERs (no notches overlap). This is confirmed in Figure~\ref{fig:05}a, where we see that compared to the SFG1s all the galaxies in the other activity groups have a higher velocity dispersion (all the confidence intervals are positive). The same is true when comparing the SFG2s to the other activity groups (not including the SFG1s), although the differences become smaller. The positive differences decrease even more when the TO1s are compared with the TO2s, the Sy2s and LINERs, or when the TO2s are compared with the Sy2s and LINERs. Finally, the Sy2s are confirmed to have significantly lower velocity dispersions than the LINERs.

In Figure~\ref{fig:04}b the box-whisker plots for the VMA show the general trend for the VMA to increase in AGNs, as was already mentioned in our introduction. The statistical significance of the differences observed are confirmed by the confidence intervals presented in Figure~\ref{fig:05}b. Note that we find exactly the same distribution for the confidence intervals of the VMA and of the velocity dispersion (compare Figure~\ref{fig:05}b with Figure~\ref{fig:05}a) . This result indicates that the VMA is more sensitive to the velocity dispersion than to the projected radius, which is a parameter in the definition of the VMA that changes with the redshift. 

In Figure~\ref{fig:04}c the change in morphology of the galaxies with different activity types follows the increase in velocity dispersion. The statistical significance of all the differences are confirmed by the intervals of confidence presented in Figure~\ref{fig:05}c. The distribution of the confidence intervals for the morphology is similar, although inverted, to the distributions previously observed for the stellar velocity dispersion and VMA.  These results imply that the morphology, the velocity dispersion and VMA are physical characteristics that are tightly connected with the activity type of the galaxy. 

Contrary to the other parameters, the absolute magnitudes in B show less differences in Figure~\ref{fig:04}d, which is also confirmed in Figure~\ref{fig:05}d by the differences in means nearer to zero. All the galaxies in the other activity groups are more luminous than the SFG1s. The TO1s and LINERs are less luminous in B than the SFG2s, while the TO2s and Sy2s are more luminous. The TO2s, the Sy2s and LINERs are all slightly more luminous in B than the TO1s. Finally, the Sy2s are slightly more luminous than the TO2s, and the LINERs are less luminous than both the Sy2s and TO2s. Note that the differences observed between the absolute B magnitudes do not trace a simple sequence with the activity type. This is consistent with the fact that the absolute B magnitude is really sensitive to two parameters, increasing with the mass of the galaxy, but also increasing with the present level of star formation. 

In Figure~\ref{fig:04}e and Figure~\ref{fig:05}e we compare the stellar masses. The SFG1s have the lowest mean mass, which increases significantly in the SFG2s and the TO1s. The TO2s and the Sy2s are more massive than these previous galaxies, the Sy2s, surprisingly, being slightly less massive than the TO2s. The LINERs are the most massive galaxies in our sample. The general trend for the difference in stellar masses is similar to the trend observed previously for the difference in velocity dispersions, except that the Sy2s are less massive than the TO2s, although they have higher velocity dispersions. 

In Figure~\ref{fig:04}f and Figure~\ref{fig:05}f there is a clear sequence for the increase of the mass-luminosity ratio with the activity type, similar to the sequences for the velocity dispersion, the VMA and the morphology. This suggests a gradual decrease in star formation as the mass of the bulge and the concentration of the mass in the center of the galaxies increase, and as the galaxies change toward earlier morphological types. 

\section{Method: the star formation history of galaxies}
\label{sec:method}

The goal of the stellar population synthesis method is to decompose, in an optimal way, the spectrum of a galaxy into simple stellar populations \citep[SSP; see explanations in][]{bruzual03}. The star formation history (SFH) is thus obtained by tracing the various components of the SSP as a function of the time covered by the grid of stellar populations used in the decomposition. 

In our study we used the population synthesis code \begin{scriptsize}STARLIGHT\end{scriptsize} \citep{CidFernandes05,Mateus06,Asari07}, which fits to the SDSS spectrum of each galaxy an optimal template spectrum. The method for determining the SFH is explained in detail in \citet{Plauchu-Frayn12}. For our decomposition we utilized a combination of $N_\star = 150$ SSP from the evolutionary synthesis models of \citet{bruzual03}, that cover a range in stellar ages from $10^5$ yrs to 20 Gyrs. This time range (including theoretically very old stars) was chosen as it allows to fit the red continuum in the oldest galaxies more easily, with a minimum of physical constraints, and it increases in general the resolution power of the synthesis code. 

From the fitted spectrum we derived the light-weighted average age, $\langle {\rm log}\; t_{\star}\rangle_{\rm L}$, and the mass-weighted average age, $\langle {\rm log}\; t_{\star}\rangle_{\rm M}$, which are equal respectively to:

\begin{equation} \langle {\rm log}\; t_{\star}\rangle_{\rm L}=\sum^N_{j=1} x_{\rm\,j}\, {\rm log}\; t_{\rm\,j}
\;\;\; {\rm and} \;\;\; \langle {\rm log}\; t_{\star}\rangle_{\rm M}=\sum^N_{j=1} \mu_{\rm\,j}\, {\rm log}\; t_{\rm\,j}
\end{equation} 

\noindent where $x_{\rm j}\; (j=1,...,N_\star)$  are the fractional contributions to the model flux weighted by light of the SSP with ages $t_j$, and $\mu_{\rm j}$ are the mass fractional contributions obtained by using the model mass-to-light ratios at the normalization $\lambda_0 = 4550$\ \AA\  (the spectra being normalized to the median flux of the continuum within the wavelength window 4530~\AA\ to 4580~\AA). The SFH is thus obtained by tracing the population vector $\mu_{\rm j}$ (or $x_{\rm j}$) as a function of $t_j$, with uncertainties of the order of $\sim 0.03$ dex \citep{Plauchu-Frayn12}. Using a smooth version of $\mu_{\rm j}$, the SFH was transformed into the star formation rate function, SFR($t_\star$), as explained in section 4.2 of \citet[][]{Asari07}. 

\begin{figure}
\includegraphics[width=\columnwidth]{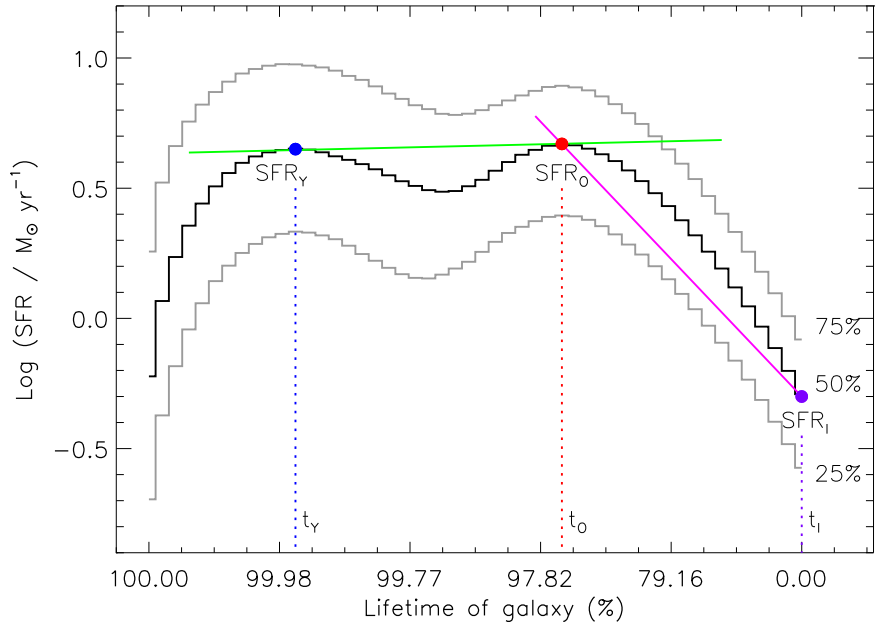}
\caption{Example of a SFH for a sub-sample of SFGs (late-type spirals). The SFH shows the variation of the median SFR($t_\star$)  over the whole lifetime of the galaxies. The two other SFHs correspond to the 25th and 75th percentiles. The parametrization of this curve is explained in the text.}
\label{fig:06}
\end{figure}

One example of a SFH in the form of the median SFR($t_\star$) for a sub-sample of SFGs (late-type spiral galaxies) is shown in Figure~\ref{fig:06}. Note that since the timescale for the SFH only reflects the theoretical range in ages covered by the grid of stellar populations used to build the SSP, we present our results for the SFH on a normalized timescale (normalized over the range $10^5$ yrs to 20 Gyrs). Our SFH shows consequently the variation of the SFR over the lifetime of a galaxy since its formation, irrespective of any cosmological model. 

For our analysis we separated the SFH in two main relevant periods:  the old period (O) covers 99\% of the lifetime of the galaxy, while the young period (Y) covers the present time, which corresponds to only the last 1\%. Note that the definition of these periods is not arbitrary, but stems from the results of our stellar population analysis: the SFH of a galaxy generally shows two peaks in SFR, one associated with the oldest stellar populations, the other with the youngest ones.  We can consequently parametrize the whole SFH using two pairs of numbers, (SFR$_{\rm Y}$,t$_{\rm Y}$) and (SFR$_{\rm O}$,t$_{\rm O}$), where each pair represents the maximum in SFR within the period, as defined above, and its relative time of occurrence (c.f. Figure~\ref{fig:06}). 

Using the above two pairs and the value of the SFR at $t = 0$ we can calculate two slopes, $S_I$,  which describes how fast the galaxy reached its maximum SFR in the past, and $S_F$, which characterizes the variation of the SFR from the past to the present. 

The mathematical definitions for the two slopes, normalized to the timescale of the stellar population grids used in our study, are the following:

\begin{equation} 
S_I = 10^{10} \times \Bigg(\frac{\Delta SFR}{\Delta t}\Bigg)_I = \frac{{\rm SFR}_{\rm O}-{\rm SFR}_{\rm I}}{{\rm t}_{\rm O}-0} 
\end{equation} 

and 

\begin{equation} 
S_ F = 10^{10} \times \Bigg(\frac{\Delta SFR}{\Delta t}\Bigg)_F = \frac{{\rm SFR}_{\rm Y}-{\rm SFR}_{\rm O}}{{\rm t}_{\rm Y}-{\rm t}_{\rm O}} 
\end{equation} 

According to our definition, galaxies with an increasing (decreasing) star formation show a positive (negative) slope. For example, for the sub-sample of SFG galaxies in Figure~\ref{fig:06} we see that they took almost 97.8\% of their lifetime to reach a maximum SFR, the slope $S_I > 0$, and then continued to form stars at more or less the same rate, the slope $S_F \sim 0$. The constant SFR observed during the young period is typical of late-type spiral galaxies \citep{Kennicutt83,Kennicutt92}. 

\begin{figure}
\includegraphics[width=\columnwidth]{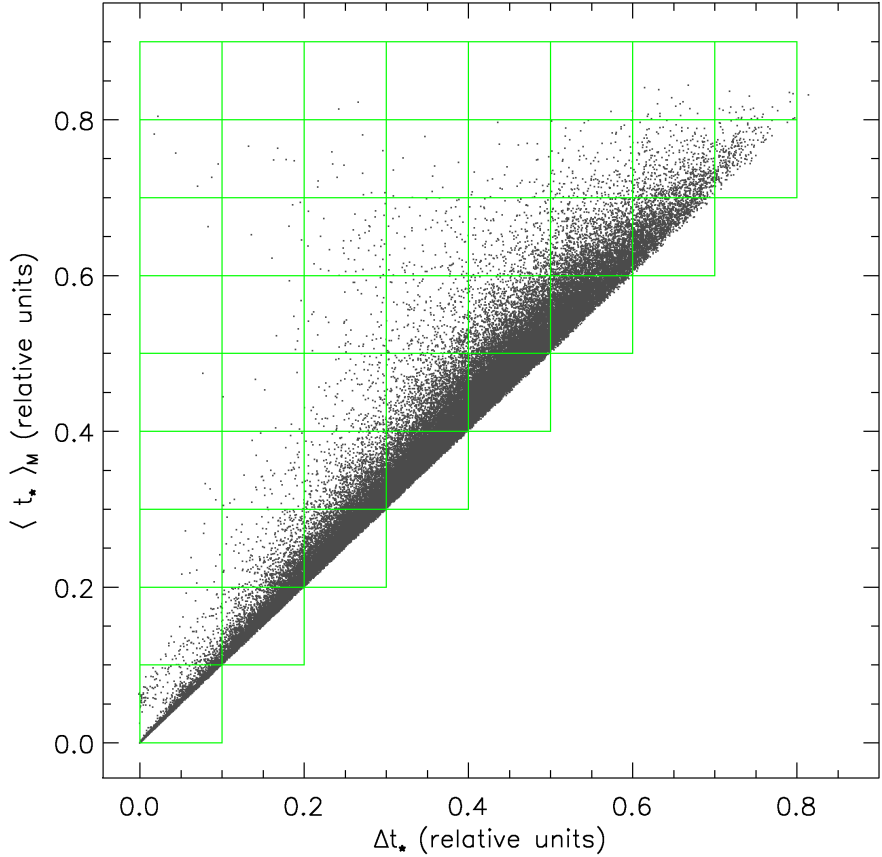}
\caption{Example of SFTS activity diagram for the SFG1s in our sample. The timescales on the two axes were normalized using the maximum age of the model SSP (the age increasing from 0 to 1). For our analysis, we separated the SFTS activity diagrams in bins (in green)  of $0.1\times0.1$ in size.}
\label{fig:07}
\end{figure}

\begin{figure}
\includegraphics[width=\columnwidth]{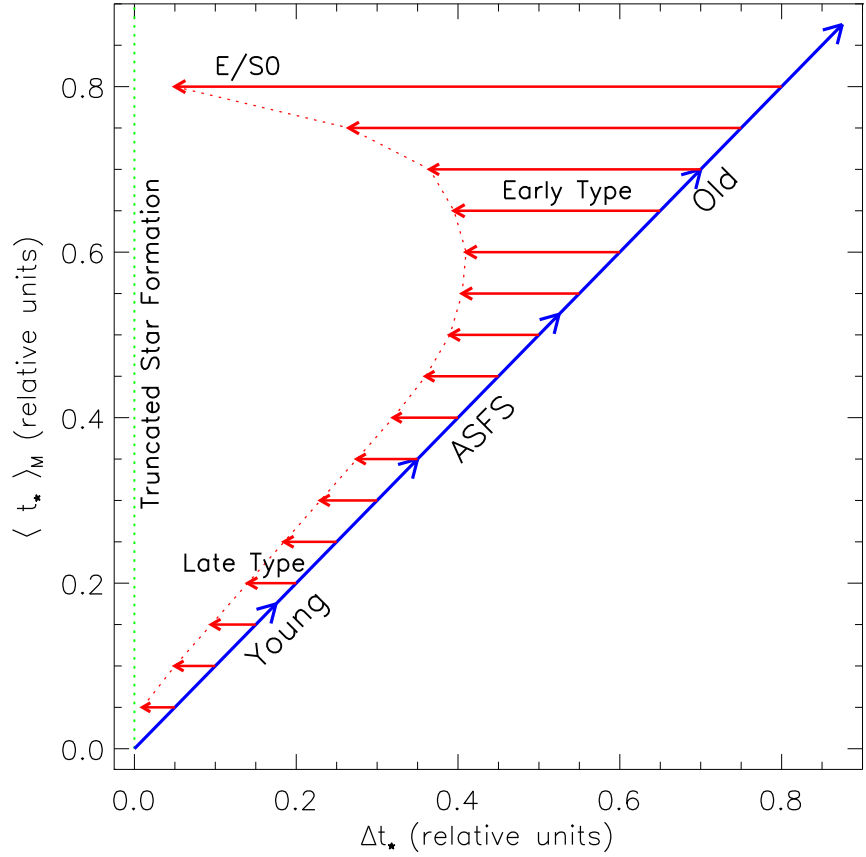}
\caption{Model SFTS activity diagram explaining what is expected for galaxies with different morphologies. The blue line (ASFS) is the active star formation sequence. The red envelope with red vectors traces the normal dispersion expected for galaxies with growing bulges. The vertical green bar to the left of the figure is the location where we expect to find candidate galaxies having suffered a truncation of star formation rate (see explanations in the text).}
\label{fig:08}
\end{figure}

Another way to analyze the SFH is by tracing the star formation timescale (SFTS) activity diagram of the galaxies \citep{Plauchu-Frayn12}. The SFTS activity diagram is obtained by tracing the mass-weighted average stellar age, $\langle t_\star \rangle_{\rm M}$,  as a function of the difference between the mass-weighted and light-weighted average stellar ages, $\Delta t_\star$. One example of a SFTS activity diagram is shown in Figure~\ref{fig:07} for the SFG1s. The principle of our analysis based on the SFTS activity diagram is the following. By definition, the two weighted average ages are sensitive to different stellar populations, $\langle {\rm log}\; t_{\star}\rangle_{\rm L}$ being sensitive to the presence of young stellar populations, because of their higher contribution in light, and $\langle {\rm log}\; t_{\star}\rangle_{\rm M}$ being sensitive to the least luminous and oldest stellar populations, which form the bulk of the mass. By analyzing the variation of $\Delta t_\star$ with time we thus obtain a picture of how fast a galaxy has formed its stars. For example, in Figure~\ref{fig:07} the SFG1s are observed to form their stars over a long period of time. As the galaxies get older, or as they change toward earlier types, $\Delta t_\star$ increases, which is due to the growing contribution of low-mass stars to the total light of the galaxies. Due to their continuous star formation activity, the SFG1s trace a narrow linear sequence in the SFTS activity diagram, $\Delta t_\star$ increasing monotonically with time. This is identified in Figure~\ref{fig:08} as the active star formation sequence (ASFS).

The full evolutionary pattern expected in the SFTS activity diagram for galaxies with different morphologies is shown in Figure~\ref{fig:08}. Any galaxy that has presently a decreasing star formation ($S_F < 0$) must be located to the left of the ASFS. It is well known that early-type galaxies form their stars more rapidly than late-type galaxies and show, consequently, decreasing star formation rates at present \citep{Sandage86,Kennicutt92,Coz11}. Therefore, as we consider more early-type galaxies in the SFTS activity diagram we expect them to be found to the left from the ASFS, and to lie further left from it as they become earlier in morphological type. Consequently, a giant elliptical galaxy, which is assumed to have formed all its stars very rapidly and very early in its life, is expected to have a small $\Delta t_\star$ and a mean old age, which would locate it at the upper left of the SFTS diagram. A spread in the duration of the star formation activity in galaxies with different morphologies and ages explains the dispersion in Figure~\ref{fig:08}. 

However, the real usefulness of the SFTS activity diagram lies in its potential to detect unusual SFHs. As we explained, any galaxy which shows presently a declining SFR must be located to the left of the ASFS. For a given morphological type, therefore, any instance of a galaxy leaving this sequence prematurely could be taken as evidence for a special, or unusual star formation process. In particular, a truncation of star formation (or quenching effect) in a galaxy would be expected to appear as a significant decrement of $\Delta t_\star$ at an abnormally young age. Such galaxy would thus be expected to be found to the extreme left of the SFTS activity diagram, anywhere, depending on its age, on the vertical green bar identified as truncated star formation. 

In principle, a thorough study of the SFH and SFTS activity diagram should reveal evidence of truncation of star formation in AGNs due to the SMBH feedback.  According to this model, we expect an AGN to show a very rapid increase in star formation, followed, after its maximum SFR in the past is reached, by a rapid decrease of star formation rate. That is, we expect a SFH characterized by a very steep positive $S_I $ and a very steep negative $S_F$. In the STFS activity diagram, we would expect these AGNs to show unusually small values of $\Delta t_\star$ for their ages and morphologies. That is they should populate more frequently the left of the SFTS diagram, either at the limit of the dispersion traced by the different morphologies or extending this limit further left. Note that the fact that we have different kinds of AGNs in our sample suggests that those that are more ``AGN like'' (according to our definition these would be the Sy2s and LINERs) should also show different behaviors  (like steeper $S_I$ slopes in the SFH and larger dispersions in the SFTS activity diagram) as compared to those with less AGN like qualities (the different TOs), and even more so compared to the SFGs. 

\section{Results}
\label{sec:results}

\begin{figure}
\includegraphics[width=\columnwidth]{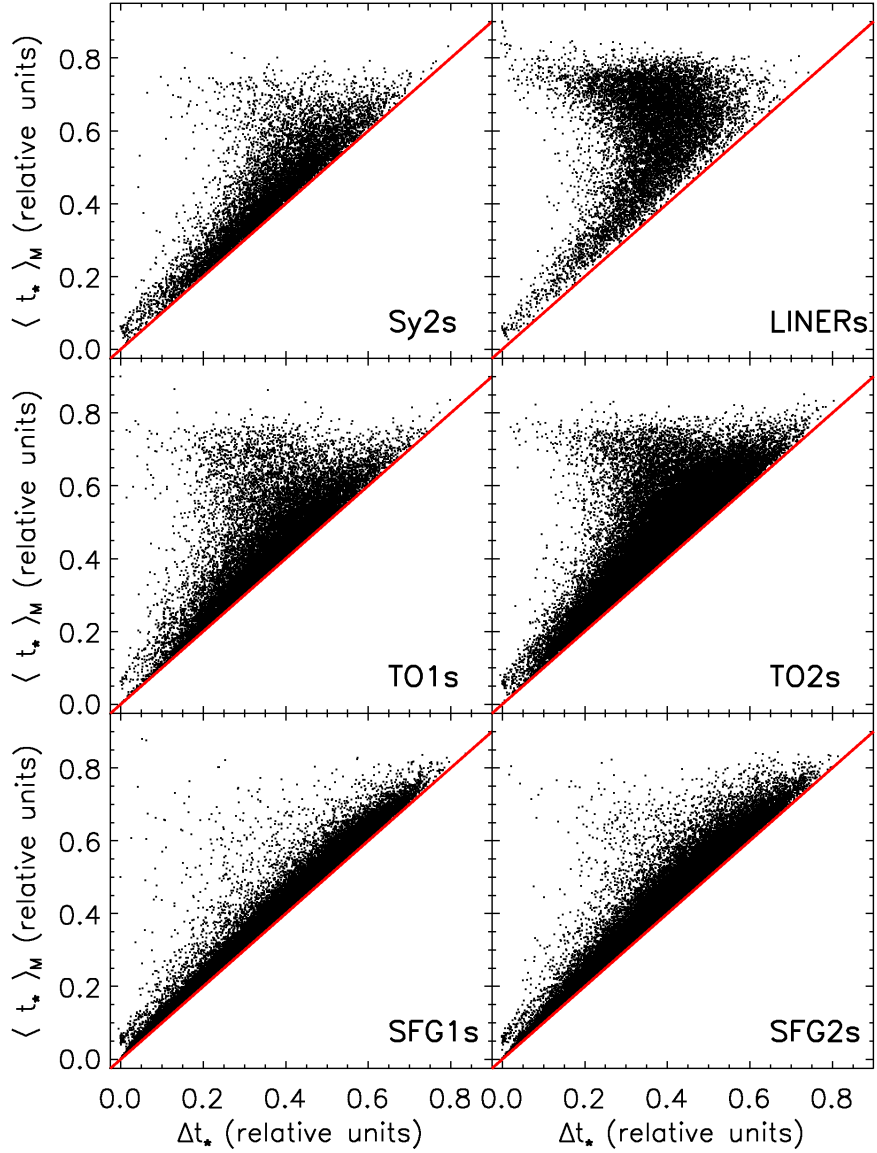}
\caption{SFTS activity diagrams for all the NELGs in our sample, separated by activity groups.}
\label{fig:09}
\end{figure}

\begin{figure}
\includegraphics[width=\columnwidth]{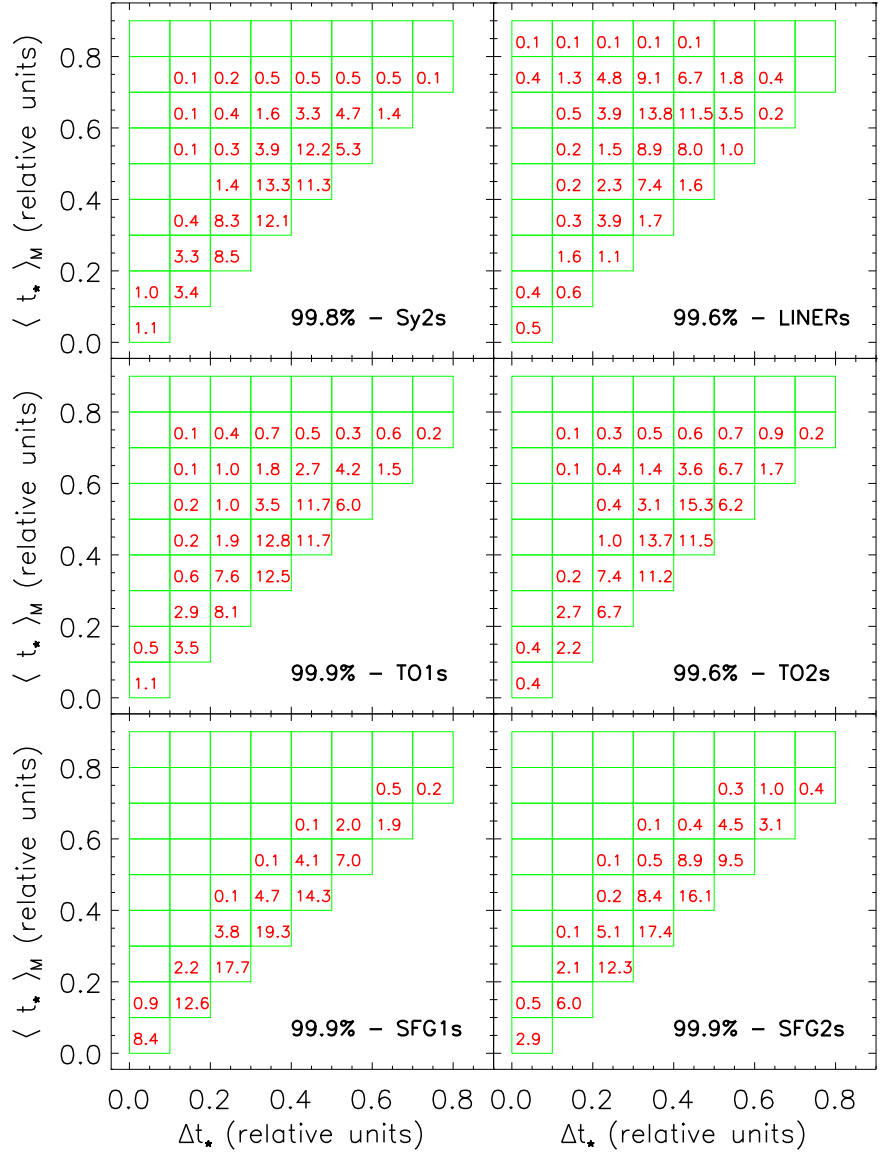}
\caption{SFTS activity diagrams showing the frequencies of galaxies in the different age bins. Bins with values below 0.1\% are not counted, which explains why the sums do not reach 100\%.}
\label{fig:10}
\end{figure}

\begin{figure*}
\includegraphics[width=\columnwidth]{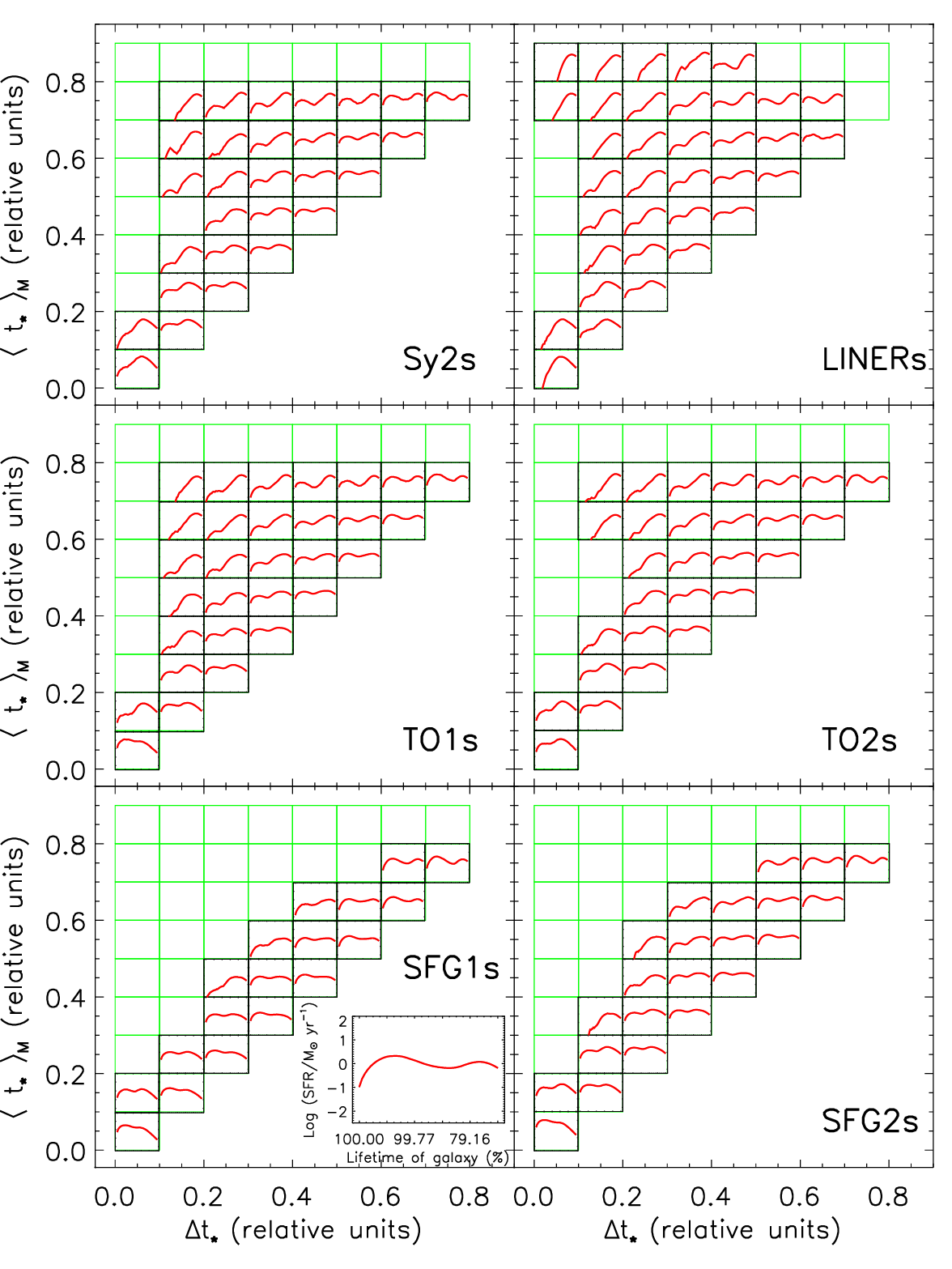}
\caption{The SFR($t_\star$)  functions as evaluated in the different bins defined in the SFTS activity diagrams. The scale of the SFR($t_\star$) function (the same in all the diagrams) is shown in the inset in the panel of the SFG1s.}
\label{fig:11}
\end{figure*}

We start our analysis by presenting in Figure~\ref{fig:09} the SFTS activity diagrams of the galaxies separated into six nuclear activity groups. In this diagram, the SFG2s trace a similar sequence as the SFG1s, only with a slightly larger dispersion. Note that at any age we see SFGs with an unusual position in the SFTS activity diagram, but this becomes less obvious in the SFG2 sample, and in both cases there are only a small number of galaxies. Curiously, but this is complicated by the variations in morphology, possible cases of unusual SFHs seem even less obvious in the TOs and AGNs than in the SFGs.  

In general, the bulk of the galaxies in the SFTS activity diagrams of the two SFG groups are located very close to the ASFS at any age. The trend shown in the SFTS  activity diagram is for the SFG2s to have higher values of $\Delta t_\star$ and $\langle t_\star \rangle_{\rm M}$ than the SFG1s. This is better observed in Figure~\ref{fig:10} where we present the percentage of galaxies in the different bins. The cumulative distribution for the SFG1s along the first two diagonals forming the ASFS is 21.9\% for $\langle t_\star \rangle_{\rm M}< 0.2$, 64.9\% for $\langle t_\star \rangle_{\rm M}< 0.4$, 95.0\% for $\langle t_\star \rangle_{\rm M} <0.6$ and 99.6\% for $\langle t_\star \rangle_{\rm M} <0.8$. The corresponding cumulative distribution for the SFG2s is 4\%, 40.9\%, 83.8\% and 92.8\%. The stellar populations are obviously getting older in the SFG2s and a slightly higher number of the galaxies (1.7\% compared to only 0.3\% for the SFG1s) are moving away from the ASFS. This behavior corresponds to a decrease in recent star formation away of the ASFS, as is observed in Figure~\ref{fig:11}, where the median SFR($t_\star$) functions for the galaxies in the different age bins of the SFTS activity diagrams are presented. The variation of the SFH follows the change in morphology and VMA (or velocity dispersion, c.f. Table~\ref{tab:01}), the SFG2s being found in slightly more early-type spirals with higher VMAs than the SFG1s.

In Figure~\ref{fig:09} the TOs show significant differences in their SFTS activity diagrams compared to the SFGs. In Figure~\ref{fig:10} the cumulative distributions for the TO1s and TO2s on the first two diagonals drop to 5.1\% and 3.0\%, respectively, for $\langle t_\star \rangle_{\rm M}< 0.2$, 36.2\% and 31.0\% for $\langle t_\star \rangle_{\rm M}< 0.4$, 78.4\% and 77.7\% for $\langle t_\star \rangle_{\rm M}< 0.6$ and 84.9\% and 87.2\% for $\langle t_\star \rangle_{\rm M}< 0.8$ and. The fraction of galaxies away from the ASFS is also increasing, reaching 15.0\% and 12.4\% for the TO1s and TO2s respectively.  In Figure~\ref{fig:11} we verify that the farther away a galaxy is from the ASFS the more pronounced is its decrement in recent star formation. Again, this variation of the SFH follows the change in morphology and VMA, the TOs residing in galaxies with earlier types and higher VMAs than the SFGs.

In Figure~\ref{fig:09} we can also establish that the LINERs are composed of galaxies that are mostly old and deficient in recent star formation. In Figure~\ref{fig:10} the cumulative distribution along the first two diagonals forming the ASFS is 1.5\%, 9.8\%, 27.8\%, reaching only 33.7\% for $\langle t_\star \rangle_{\rm M} <0.8$. The fraction of LINERS away from the ASFS increases significantly to 65.9\%, with 52.5\% above $\langle t_\star \rangle_{\rm M} = 0.6$. In Figure~\ref{fig:11} for the LINERs, we observe that on the first two diagonals defining the ASFS in the SFTS activity diagram, the median SFR($t_\star$) functions are already decreasing at recent times. Here too, the variation of SFH follows the general change toward earlier morphological types and higher VMAs. 

Curiously, we do not observe in Figure~\ref{fig:09} a similar aging trend for the Sy2s as for the LINERs. Despite the changes toward earlier types in morphology and higher VMAs (c.f. Table~\ref{tab:01}), the Sy2s look unusually ``young'', showing high SFRs at present. In Figure~\ref{fig:10} the cumulative distribution along the first two diagonals is 5.5\%, 37.7\%, 79.8\%  and  86.5\%. The number of Sy2s away from the ASFS is consequently quite small, only 11.5\%. In terms of star formation, the Sy2s look similar to the TOs, and show comparable decrements in star formation away from the ASFS in Figure~\ref{fig:11}. Therefore, the Sy2s appear to show an ``excess'' of star formation for their morphology and VMA. 

\begin{figure}
\includegraphics[width=\columnwidth]{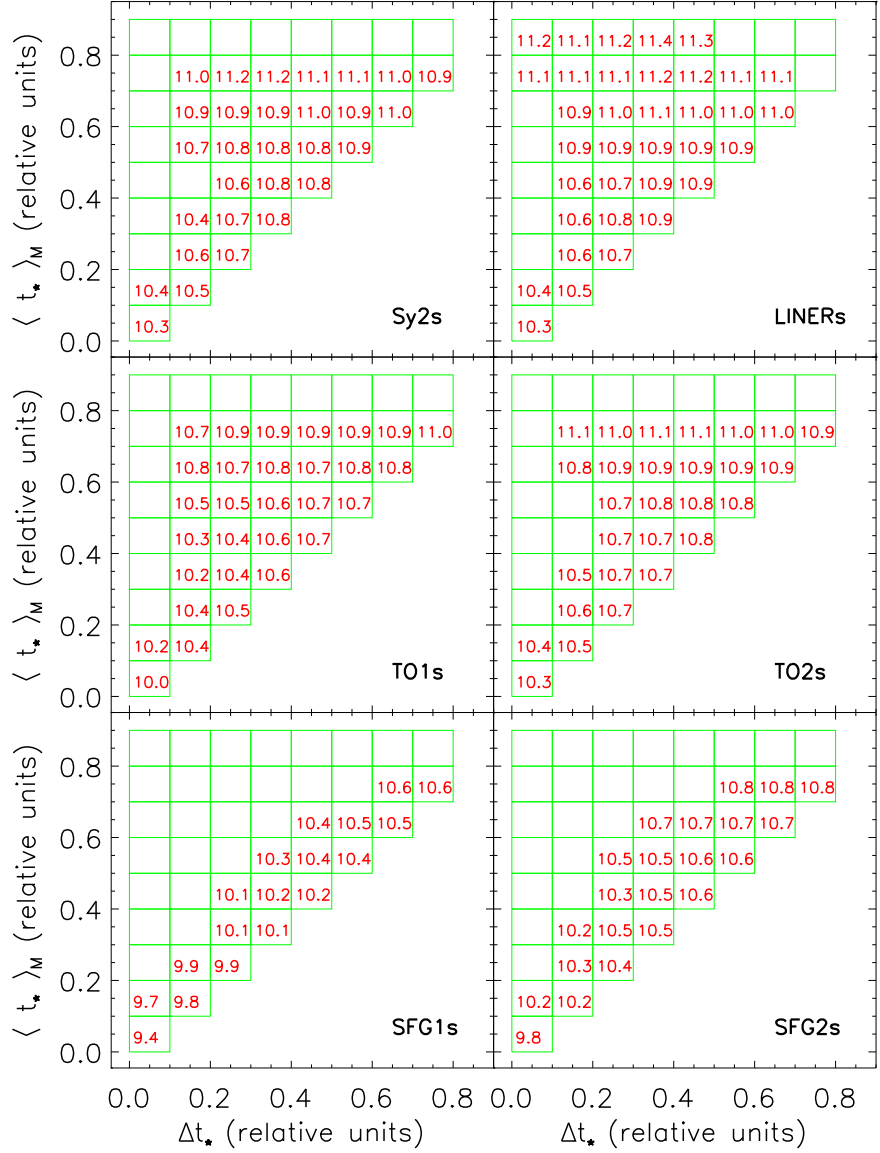}
\caption{Variations of the mean stellar mass in the SFTS activity diagrams of galaxies with different activity types. In each bin we give the logarithm of the mass in solar unit.}
\label{fig:12}
\end{figure}

\begin{figure}
\includegraphics[width=\columnwidth]{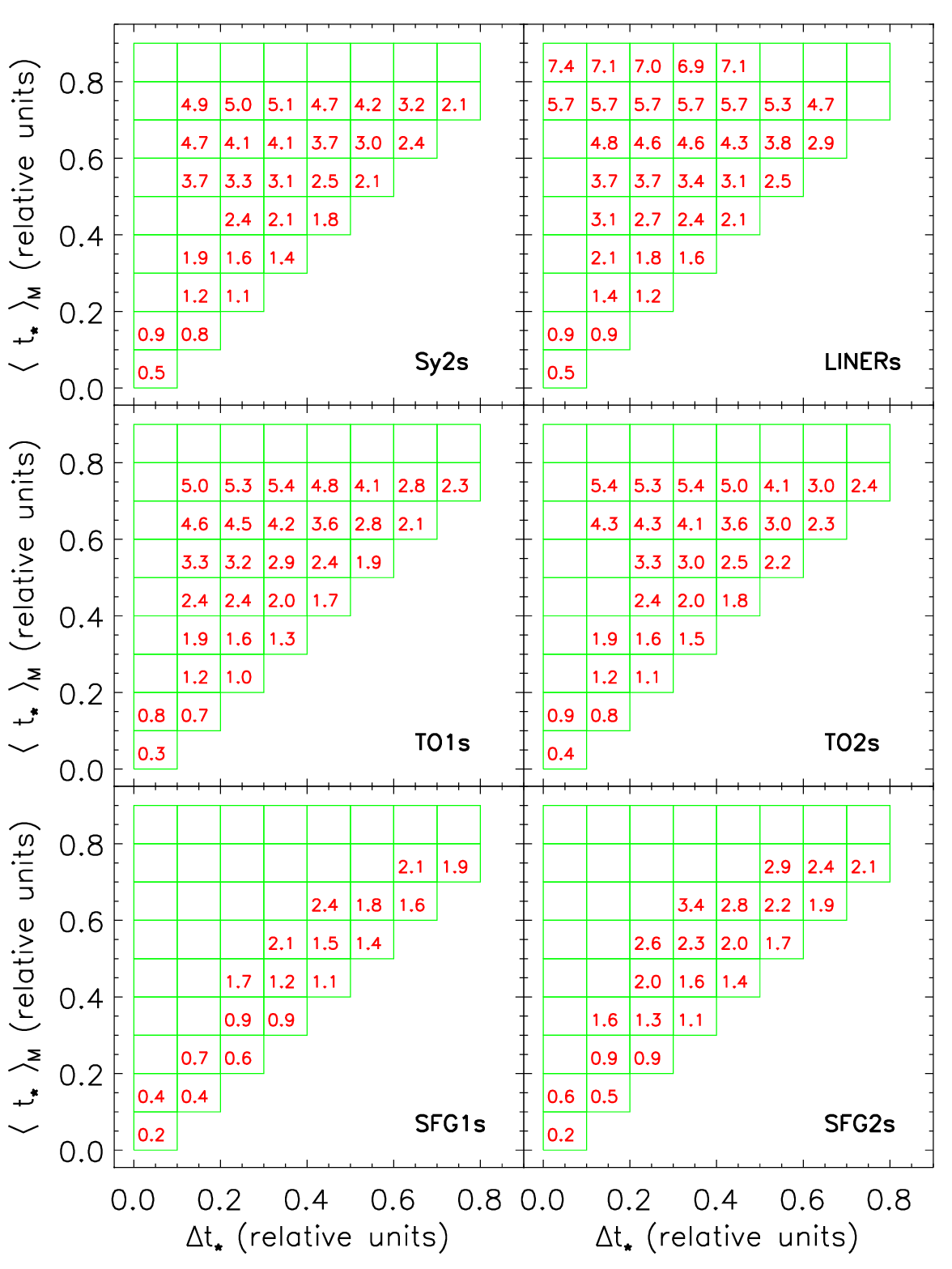}
\caption{Variations of the mean mass-luminosity in the SFTS activity diagrams of galaxies with different activity types.}
\label{fig:13}
\end{figure}

In Figure~\ref{fig:12} we show how the mean mass of the galaxies in each bin of the SFTS activity diagram varies with the activity type. As already noted, in general the mass increases from the SFGs to the TOs and Sy2s,  to the LINERs. However, Figure~\ref{fig:12} also reveals that the mass systematically increases with $\langle t_\star \rangle_{\rm M}$ regardless of the activity type, such that the galaxies that have the oldest stellar populations are also the most massive ones. Also, at any mean age $\langle t_\star \rangle_{\rm M}$ we see a trend for the mass to decrease away from the ASFS as $\Delta t_\star$ decreases. Independently of the activity type, therefore, the more massive galaxies at any age tend to extend their star formation activity over relatively longer periods of time than the less massive ones. This is confirmed by the variations of the mass-luminosity ratio in the SFTS activity diagram presented in Figure~\ref{fig:13}. The mass-luminosity ratio increases with the mean age $\langle t_\star \rangle_{\rm M}$, and increases even more at any $\langle t_\star \rangle_{\rm M}$ away from the ASFS as $\Delta t_\star$ decreases. In fact, only for low values of  $\Delta t_\star$ with high $\langle t_\star \rangle_{\rm M}$ do we observe values of mass-luminosity normally consistent with the morphology of the galaxies. Therefore, it seems that in all the NELGs in our sample the normal phases of star formation cover relatively long periods of time. 

\begin{figure}
\includegraphics[width=\columnwidth]{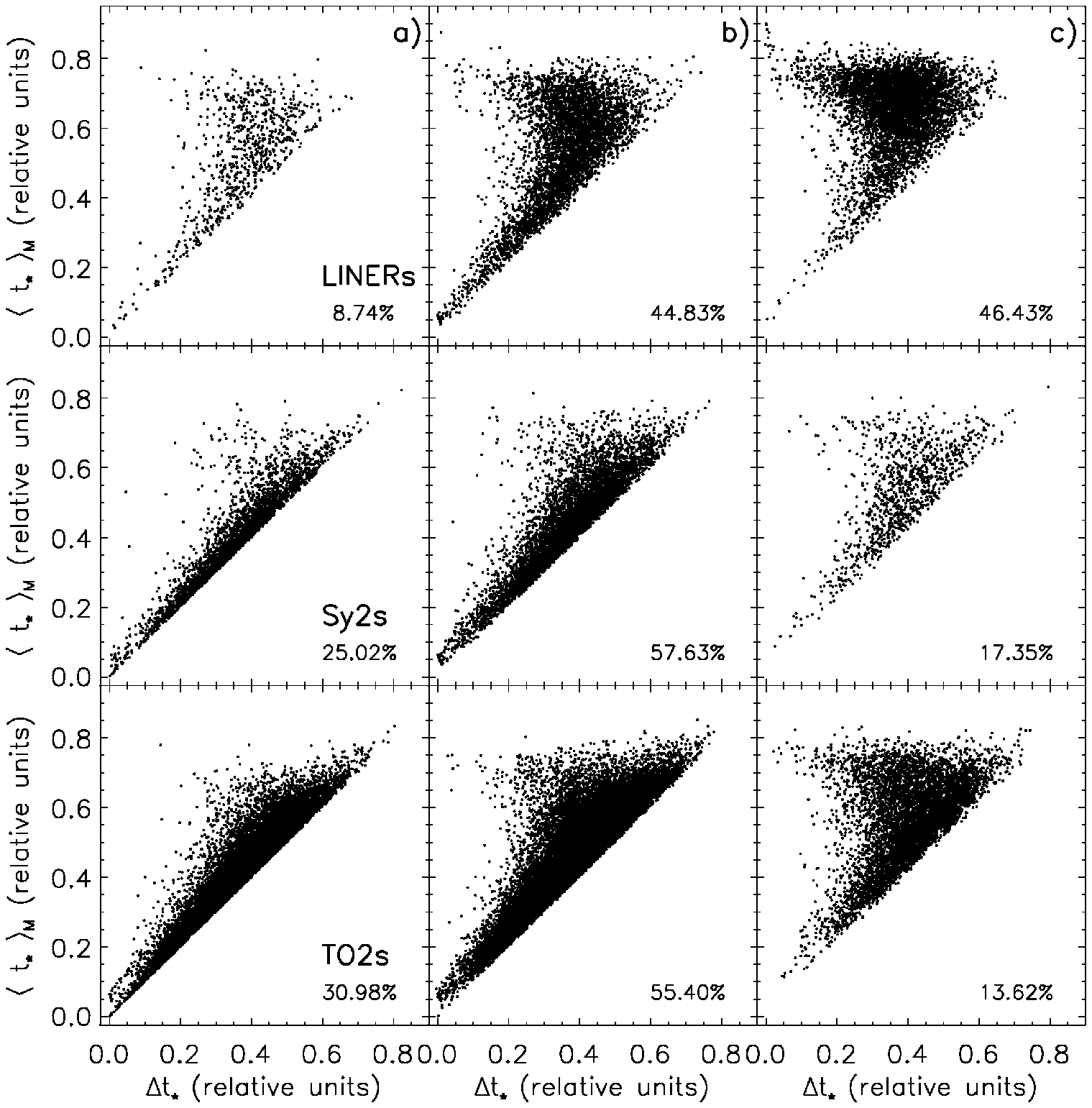}
\caption{The variation of morphologies in the SFTS activity diagrams for the LINERs, Sy2s and TO2s: a) Late (Sbc to Im), b) Intermediate (Sa to Sb), and c) Early (E to S0/a).} 
\label{fig:14}
\end{figure}

\begin{figure}
\includegraphics[width=\columnwidth]{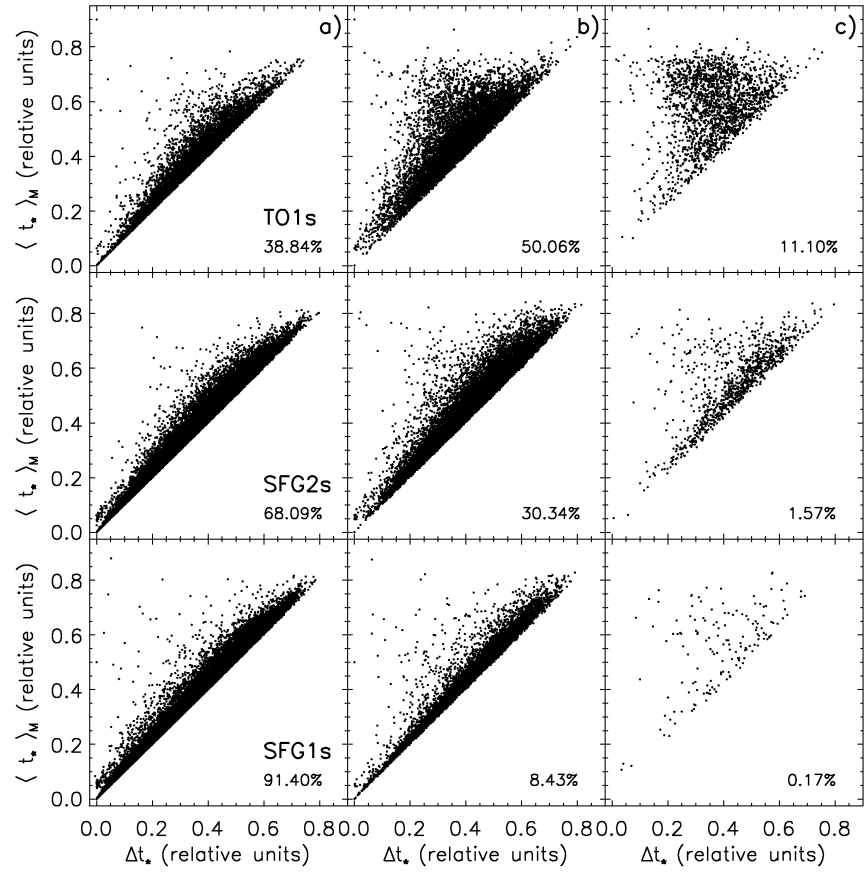}
\caption{The variation of morphologies in the SFTS activity diagrams for the TO1s, SFG2s and SFG1s: a) Late (Sbc to Im), b) Intermediate (Sa to Sb), and c) Early (E to S0/a).} 
\label{fig:15}
\end{figure}

\begin{figure}
\includegraphics[width=\columnwidth]{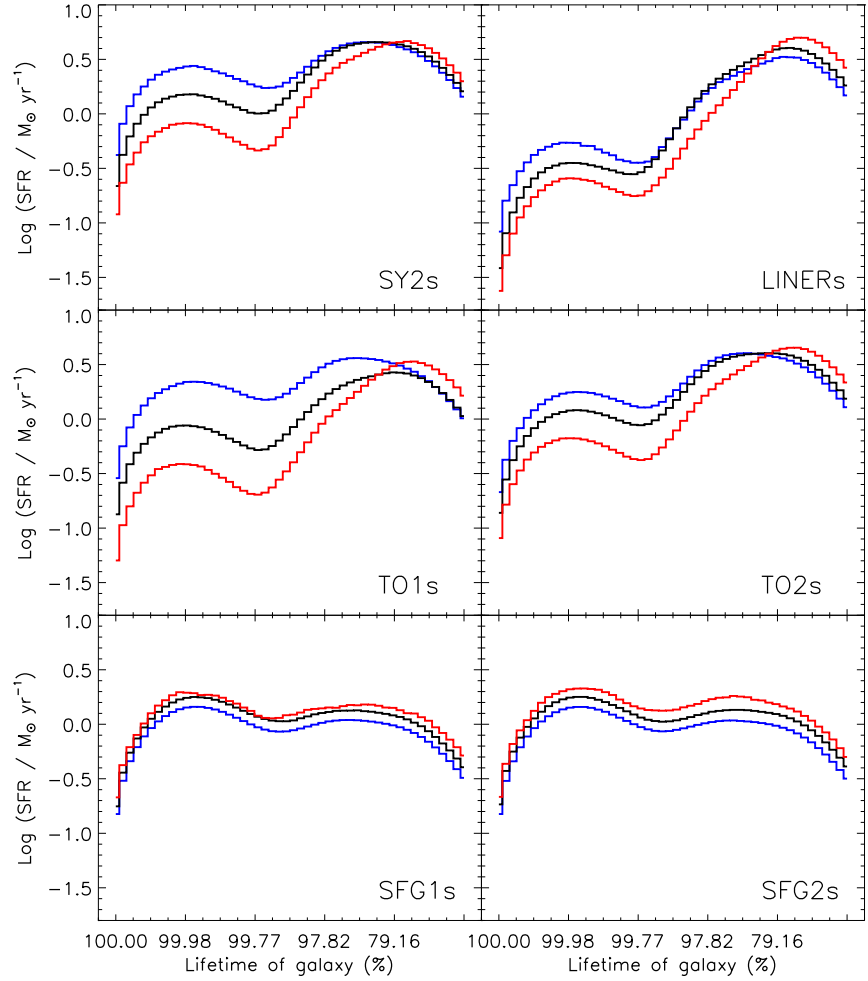}
\caption{The variations of SFH in the SFTS activity diagrams for the NELGs with different activity types, separated in three sub-samples in morphology: in blue Late, in black Intermediate, and in red Early.} 
\label{fig:16}
\end{figure}

\begin{table*}[!t]\centering
  \small
  \newcommand{\DS}{\hspace{6\tabcolsep}} 
 \begin{changemargin}{-1.0cm}{-1.0cm}
    \caption{Characteristic SFH parameters and slopes in the most populated bins of the SFTS activity diagrams} \label{tab:03}
    \setlength{\tabnotewidth}{0.95\linewidth}
    \setlength{\tabcolsep}{0.4\tabcolsep} \tablecols{10}
    \begin{tabular}{lccccccccc}
      \toprule
(1)&(2)&\DS (3)&(4)*&(5)&(6)*&(7)*&(8)*&(9)&(10)\\
 Act.   & freq.     &\DS  t$_Y$ & SFR$_Y$& t$_O$ & SFR$_O$ &  SFR$_I$ & VMA  &  $S_F$  & $S_I$\\
 type   &    ( \%)  &\DS   (\%) &  (M$_\odot$ yr$^{-1}$) &  (\%)   & (M$_\odot$ yr$^{-1}$) &  (M$_\odot$ yr$^{-1}$) &  (M$_\odot$)  &  (M$_\odot$ yr$^{-1}$) & (M$_\odot$ yr$^{-1}$)\\
\midrule
 LINERs & 13.8 &\DS 99.97  & $-$0.83  &  48.60  & \ \ 0.69  & \ \ 0.41  &10.22 & $-$4.75 & 2.33 \\
 Sy2s    & 13.3 &\DS 99.98  & \ \ 0.18   &  92.45  & \ \ 0.66  & \ \ 0.21  &10.09 & $-$1.93 & 1.61 \\
 TO2s   & 15.3 &\DS 99.97  & $-$0.01  &  67.27  & \ \ 0.48  & \ \ 0.14  &  9.94 & $-$2.24 & 1.20 \\
 TO1s   & 12.8 &\DS 99.97  & $-$0.08  &  88.14  & \ \ 0.39  & $-$0.05  &  9.88 & $-$2.47 & 0.89 \\
 SFG2s  & 17.4 &\DS 99.96  & \ \ 0.38  &  95.19  & \ \ 0.44  & $-$0.13  &  9.87 & $-$1.58 & 1.06 \\
 SFG1s  & 19.3 &\DS 99.97  & \ \ 0.17  &  96.94  & $-$0.05  & $-$0.64  &  9.53 & $+$1.41 & 0.34 \\
      \midrule
 LINERs & 51.3 &\DS 99.98  & $-$0.60  &  48.60  & \ \ 0.69 & \ \ 0.40 & 10.21 & $-$4.64 & 2.39 \\
 Sy2s    & 48.9 &\DS 99.98  & \ \ 0.30  &  90.54  & \ \ 0.66 & \ \ 0.24 & 10.09 & $-$1.49 & 1.57 \\
 TO2s   & 51.7 &\DS 99.98  & \ \ 0.17  &  85.14  & \ \ 0.59 & \ \ 0.18 & 10.02 & $-$2.41 & 1.41 \\
 TO1s   & 48.7 &\DS 99.98  & \ \ 0.15  &  88.14  & \ \ 0.44 & \ \ 0.02 &  9.93 & $-$2.37 & 0.97 \\
 SFG2s  & 55.3 &\DS 99.97  & \ \ 0.38  &  95.19  & \ \ 0.42 & $-$0.16  &  9.86 & $-$1.63 & 1.02 \\
 SFG1s  & 51.3 &\DS 99.97  & \ \ 0.18  &  96.93  & $-$0.05  & $-$0.63  &  9.54 & $+$1.46 & 0.34 \\
      \midrule
 LINERs & 74.1 &\DS 99.98  & $-$0.48  &  58.98  & \ \ 0.68 & \ \ 0.40 & 10.20 & $-$4.58 & 2.19 \\
 Sy2s    & 75.7 &\DS 99.98  & \ \ 0.31   &  90.53  & \ \ 0.67 & \ \ 0.22 & 10.08 & $-$1.52 & 1.48 \\
 TO2s   & 72.0 &\DS 99.98  & \ \ 0.19  &  85.14  & \ \ 0.59 & \ \ 0.16 & 10.02 & $-$2.53 & 1.45 \\
 TO1s   & 74.6 &\DS 99.98  & \ \ 0.17  &  88.14  & \ \ 0.44 & \ \ 0.01 &  9.93 & $-$2.16 & 0.99 \\
 SFG2s  & 72.6 &\DS 99.98  & \ \ 0.30  &  93.98  & \ \ 0.33 & $-$0.17  &  9.82 & $-$1.84 & 0.78 \\
 SFG1s  & 72.3 &\DS 99.98  & \ \ 0.18  &  96.16  & $-$0.04  & $-$0.63  &  9.55 & $+$1.47 & 0.35 \\
      \bottomrule
    \end{tabular}\\
{\small  In columns followed by an asterisk, the logarithm of the parameter is listed.}
\end{changemargin}
\end{table*}

In Figure~\ref{fig:14} for the AGNs and TO2s and in Figure~\ref{fig:15} for the SFGs and TO1s we present the SFTS activity diagrams of the galaxies separated into three sub-classes of morphologies:  Late (Sbc to Im), Intermediate (Sa to Sb) and Early (E to S0/a). Note that we do not find a clear correlation between the morphology and the position of the galaxies on the SFTS activity diagram. Only the SFGs show such a high level of correlation, but this is because they are formed mostly of Late galaxies. When the morphology changes toward earlier types, the stellar formation activity becomes dispersed over differently long periods of time, generally increasing with the mass of the galaxies as observed in Figure~\ref{fig:12}. This common behavior, which is independent of the activity type, suggests that no galaxies in our sample present evidence of a truncation of their star formation.   

In Figure~\ref{fig:16} we have calculated the median SFR($t_\star$) functions for the galaxies with different activity types separated in the three sub-samples in morphologies as was defined in Figure~\ref{fig:14} and Figure~\ref{fig:15}. No variation is observed for the two SFGs. Note that the SFRs are higher at the present time than in the past, which means that $S_F >0$ in these galaxies.  

For the Early LINERs, Sy2s and TOs, we see that their SFHs are quite similar. The SFR in the past was much higher than at present. We note also that as the morphology becomes later in type the maxima of star formation in the past move toward higher fractions of the lifetime of the galaxies and the present SFR increases significantly. However, these variations are much less important in the LINERs and TO2s than in the TO1s or Sy2s.

\begin{figure}
\includegraphics[width=\columnwidth]{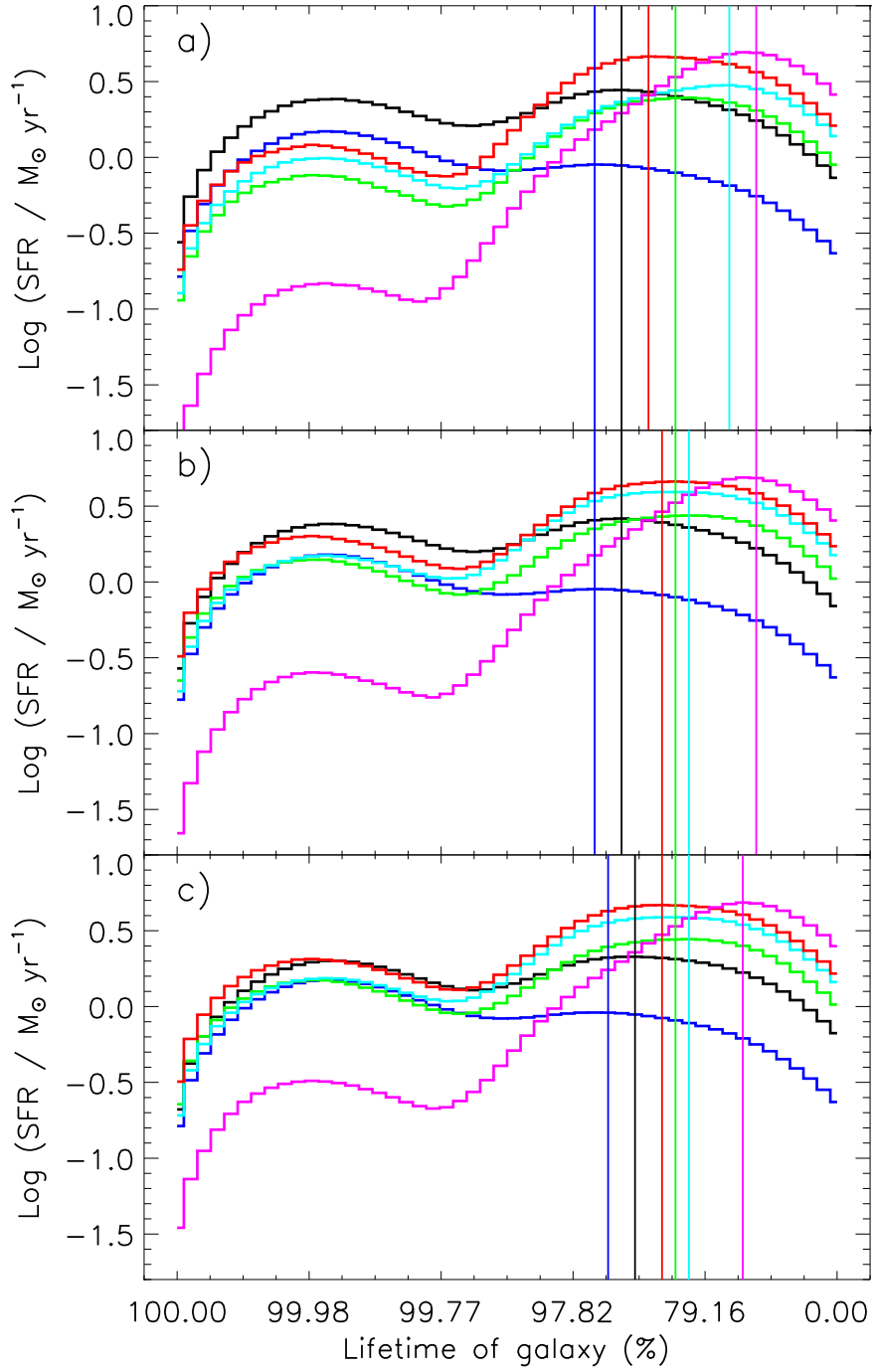}
\caption{Median SFR($t_\star$) functions as measured in the bins of the SFTS activity diagrams having highest frequencies. The number of galaxies counted increases from a) to c). The color code is the same as in Figure~\ref{fig:02}. The vertical lines identify the time when the SFR was at its highest in the past (t$_{O}$).} 
\label{fig:17}
\end{figure}

\begin{figure}
\includegraphics[width=\columnwidth]{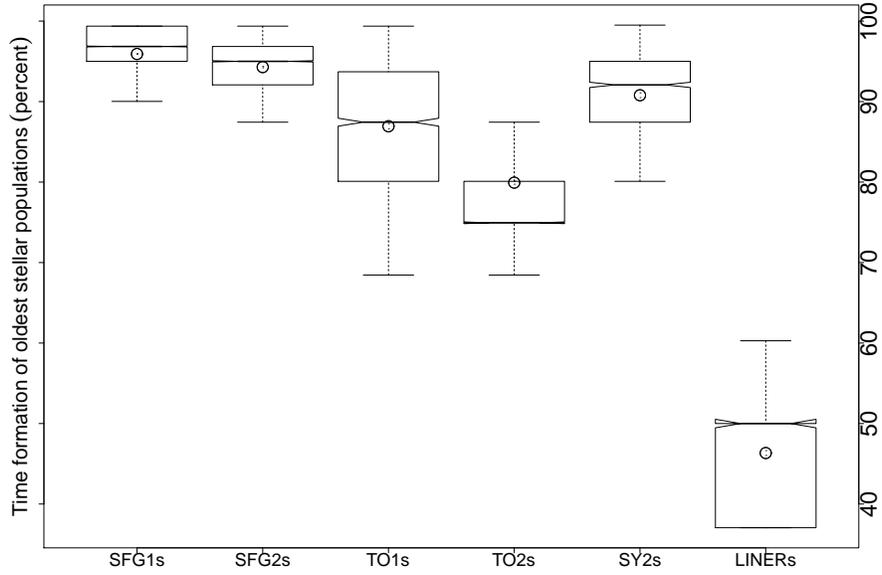}
\caption{Box-whisker plots comparing t$_{O}$.} 
\label{fig:18}
\end{figure}

\begin{figure}
\includegraphics[width=\columnwidth]{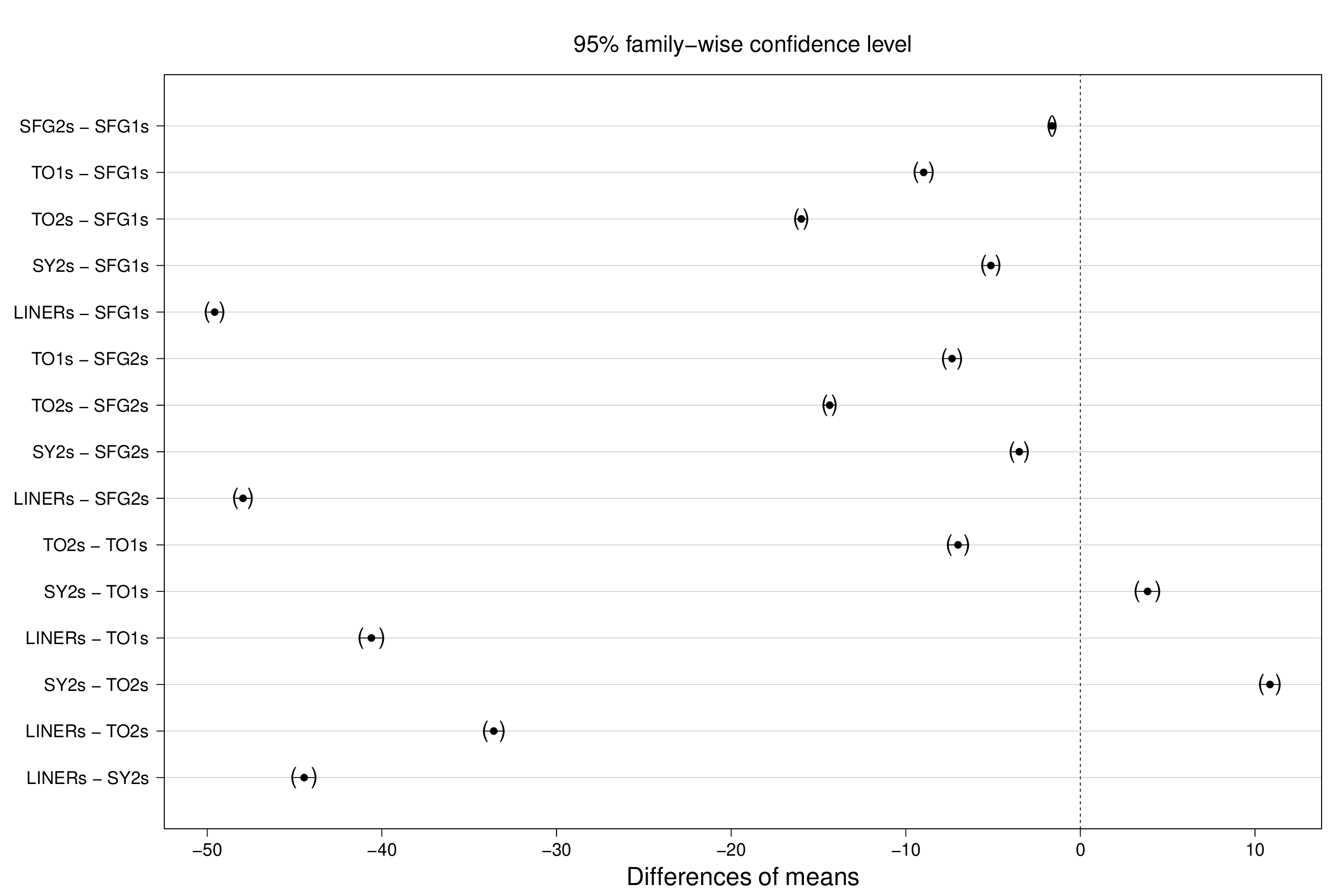}
\caption{Simultaneous confidence intervals for t$_O$.}
\label{fig:19}
\end{figure}

It is clear from the above analysis that what determines the normal SFH of galaxies with different activity types is their predominant morphologies: Early for the LINERs, Intermediate for the Sy2s and TOs and Late for the SFGs. This is what we observe, consequently, in Figure~\ref{fig:17}, where we present the median SFR($t_\star$) functions for galaxies with different activity types as calculated within the bins of the SFTS activity diagrams (Figure~\ref{fig:10}) that have the highest galaxy frequencies. The cumulative frequencies and the median values characterizing the SFR($t_\star$) functions are reported in Table~\ref{tab:03}. The upper diagram (Figure~\ref{fig:17}a) shows the median SFR($t_\star$) functions as calculated in the most populated bin for each activity group. The two lower diagrams (Figure~\ref{fig:17}b and c) include more and more bins in the calculation of the medians. 

Compared to the SFG1s, the LINERs show more intense star formation rates in the past (almost 6 times higher than the SFG1s and 2 times higher than the SFG2s). They also formed the bulk of their stars earlier than the SFGs. It took the LINERs only $\sim 49$\% of their lifetime to reach their maximum in SFR, compared to $\sim 95$\% and $\sim 97$\% for the SFG2s and SFG1s respectively. These numbers do not change much as we include more galaxies (due to small variations in morphology). 

The TOs also formed most of their stars before the SFGs. It took the TO2s $\sim 67$\% of their lifetime and the TO1s $\sim 88$\% to reach their maximum SFRs. Note however that when we increase the number of galaxies in the calculation these numbers do change significantly for the TO2s, reaching $\sim 85$\%  in Figure~\ref{fig:17}c (the value for the TO1s stays at 88.1\%), which makes them more similar to the TO1s. 

The most curious behavior, however, is for the Sy2s, which took almost the same time as the SFGs ($\sim 91\%-92$\%) to reach their maximum in SFR. Note, on the other hand, that the SFR in the past for the SY2s was as high as for the LINERs, which explains, in part, why they show an excess in star formation today. 

Examining the slopes, we find that only the LINERs experienced an important increase of star formation early in their lifetime ($S_I  > 2$) and a significant decrease after that ($S_F < -4$). However this behavior does not describe an instantaneous event, since, as we have shown, it took the LINERs $\sim 49$\% of their lifetime to reach a maximum in SFR in the past.  Although the decrement in star formation after reaching this maximum is significant, it is also by no means instantaneous, but rather gradual, with the normal phase of stellar activity, as we have shown before, extending over differently long periods of time. 

The two slopes for the Sy2s and TOs are very much similar, as expected based on the morphological similarities. The SFR decreased much more slowly in the Sy2s than the TOs, which also contribute in explaining the excess of star formation at present in these galaxies. The SFG2s follow the same trend as the TOs and Sy2s, with $S_I >0$ and $S_F<0$, differing only by their lower SFR in the past. The SFG1s, on the other hand, show the smallest $S_I$ with $S_F >0$. Once again, the difference between the SFG1s and SFG2s can be explained by the different mass and different morphologies.

The most significant differences in the SFHs is the different time in the past when the SFR was at its maximum (t$_O$). In Figure~\ref{fig:18} we have traced the box-whisker plots comparing t$_O$ in NELGs with different activity types. This diagram shows that the LINERs formed their stars much more rapidly than any of the other galaxies in our sample. The TO2s then formed their stars less rapidly than the LINERs but more rapidly than the TO1s. The SFGs and Sy2s formed their stars much more slowly. Except for the Sy2s, therefore, this sequence is well correlated with the stellar mass: the most massive galaxies formed their stars more rapidly than the less massive ones.  The confidence intervals presented in Figure~\ref{fig:19} confirm the differences observed between the activity groups at a level of confidence of 95\%.

\section{Discussion}
\label{sec:discussion}

The main goal of our study is to establish the normal mode of star formation in galaxies presenting different activity  types. By doing so we hope to be able to  determine if the SMBHs in the AGNs have had an influence on their formation, e.g. in the form of a truncation, or quenching of their star formation. Applying our working definition for the truncation of star formation we should have expected to observe in the AGNs of our sample a rapid increase of their star formation rate in the past and a sudden decrease after reaching their maximum. Also, we should have expected to observe unusual SFHs in the SFTS, under the form of an extremely short period of star formation.

 Our analysis of the SFHs reveals no such evidence in the Sy2s and TOs. In fact we found the contrary, that all these galaxies are still actively forming stars, and the Sy2s, in particular, may show an excess of star formation for their mean mass, morphology and VMA. 

The only galaxies that present some characteristics that could be consistent with our working definition of a quenching effect in AGNs are the LINERs. However, we have found that these galaxies did not form their stars in an instantaneous burst, but rather that it took them almost 50\% of their lifetime to reach their maximum in SFR. And although after that the star formation is found to decrease significantly, this decrement is rather gradual, the normal phase of stellar formation activity being spread over relatively long periods of time, which varies with the mass of the galaxies. In fact, this dispersion in time of the stellar formation activity is a common trait of all the NELGs in our sample, independent of the morphology or activity type.

Based on the above analysis, we conclude therefore that we see no evidence of a truncation of star formation, or quenching effect, in any of the NELGs in our sample, AGN or not. On the other hand, what we observe seems fully consistent with the model for the formation of galaxies in general, showing that the most massive galaxies build their stars more rapidly than the less massive ones, and that these galaxies turned out at the end to have more massive bulges. The mass and the morphology, or rather the importance of the bulge, seem, therefore, the only connections we found between the SFHs and the AGN phenomenon. 

\begin{figure}
\includegraphics[width=\columnwidth]{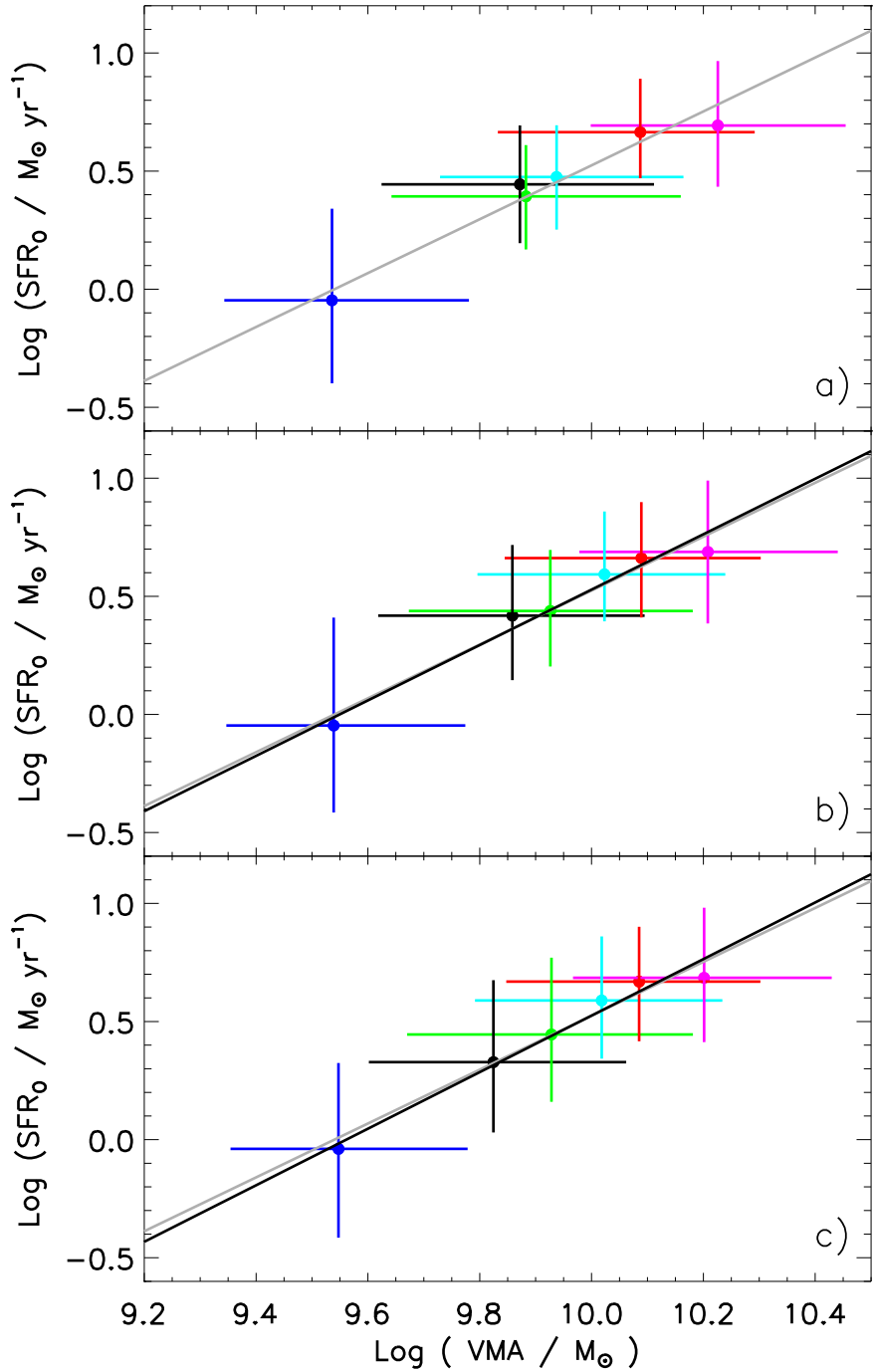}
\caption{Relation between the maximum SFR in the past and the VMA as calculated for the galaxy sub-samples of Figure~\ref{fig:17}. The color code is the same as in Figure~\ref{fig:02}. The dots correspond to the medians and the bars correspond to the 25th and 75th percentiles. The linear regression obtained in each sub-sample is shown and compared in b) and c) with the one found in a) (the lighter grey curve).}
\label{fig:20}
\end{figure}

However, in \citet{TP12a} it was shown that it is the VMA, which characterizes the concentration of mass in the center of the galaxy, that is better correlated with the activity type of the galaxies, not the morphology. That is, comparing galaxies with the same morphology, the AGN always appear in the galaxy that have the highest VMA. How can we reconcile this observation with the results for the SFHs? In Figure~\ref{fig:20} we show that the maximum SFR in the past is well correlated with the VMA  (with a coefficient of correlation of 0.97). This new correlation suggests that the formation of a SMBH at the center of galaxies happened during a phase of intense star formation, and the higher the intensity of star formation the higher the mass of the SMBH, and, consequently, the higher the chance to observe an AGN \citep{Silk81,Ebisuzaki01, Carilli01,Mouri02,PortegiesZwart04, Farrah04,Gurkan04,Mobasher04,Schweitzer06,Escala07,Rafferty11,Goto11,Cen12}. 

To understand better this relation between the VMA and maximum SFR in the past, we must look at models for the formation of the SMBHs at the center of galaxies.  For example, in the models of \citet{Ebisuzaki01} and \citet{Mouri02}, intermediate-mass black holes form first in young compact star clusters through runaway merging of massive stars. According to these models, therefore, the higher the number of clusters formed (the higher the level of star formation in the past), the higher the number of massive stars, and, consequently, the higher the mass of the black hole. Moreover, in these models the host star clusters are also predicted to sink toward the galactic nucleus through dynamical friction, which means that the process is dissipative. Writing the mechanical energy in its usual form, as the sum of the kinetic and potential energies, we would thus obtain the following relation:

\begin{equation}
\frac{dE}{dt} = \frac{dK}{dt}-\frac{d}{dt}\left| \frac{GM^2}{R}\right|  < 0
\end{equation}

\noindent where the first term on the right of Eq. 5 corresponds to the dissipation of kinetic energy and the second to an increase in potential gravitational energy. This process would naturally connect the formation of the SMBH to an increase of the gravitational potential energy at the center of the galaxy  (or binding energy of the nucleus), explaining the correlations between the AGN types and the VMA \citep{TP12a}. 

Note that the farther in the past is the maximum of star formation and the higher the probability for this maximum to be connected with the formation of the bulge (and SMBHs) through the mergers of protogalaxies. Since protogalaxies have less angular momentum, the movement of mass toward the center of the forming galaxies would thus be facilitated, which would fit our observations for the LINERs: they formed most of their stars in the distant past and have the highest VMA of all the AGNs. 

For the TOs which formed their stars more recently, and have slightly later morphological types, other mechanisms could be necessary to dissipate the rotational energy \citep[e.g.][]{Kawakatu02,Kawakatu04}. This would be consistent with their lower SFRs and lower VMAs compared to the LINERs (also implying lower masses for the SMBHs at their centers). 

 It is true that the Sy2s are more similar in morphology to the TOs than the LINERs. They seem on the other hand more similar to the LINERs than the TOs in terms of their level of star formation in the past and their VMA. The only obvious difference, therefore, seems to be how slowly, or how recently, they formed their stars. Could the differences between the LINERs and Sy2s be due solely to a difference in angular momentum? Unless we find an obvious difference in the formation process of these galaxies (for example, evidence for recent, major mergers, or a higher rate of interactions with neighboring galaxies) then the Sy2s may just be the most recent examples of massive galaxies forming a SMBH at their center.    

\section{Summary and conclusions}
\label{sec:conclusion}

Using a sample of  229618 NELGs previously studied by \citet{TP12a}, we have analyzed their SFHs in order to determine how galaxies with different nuclear activity types normally formed their stars. The results of our analysis reveal new insights on the Starburst-AGN connection and possibly also the origin of the M$_{\rm BH}$-$\sigma_*$ relation. 

We conclude that there is no evidence of an influence of the SMBH in terms of a truncation of star formation in any of the NELGs in our sample (AGN or not). On the contrary, active star formation activity is quite common, and the most massive galaxies, independent of their activity type and morphology, show the longest star formation episodes. We have also found that the Sy2s show at present an excess of star formation considering their mean mass, morphology and VMA.

In general, the more massive galaxies formed the bulk of their stars more rapidly than the less massive ones. We have also found that the variation of the SFH with the activity type is explained by the masses of the galaxies and by the importance of the bulge. We conclude, therefore, that the mass and the morphology, or rather the importance of the bulge, seem like the only connections we found between the SFH and the AGN phenomenon.

Consistent with the above conclusion we have found that the maximum in star formation rate in the past increases with the VMA. This correlation can be explained assuming galaxies form a SMBH at their centers through highly dissipative processes related with the formation of their bulges. 

>From our analysis we conclude, therefore, that the M$_{\rm BH}$-$\sigma_*$ relation is most probably a consequence of the fact that the formation of the bulge and of the SMBH at its center are simultaneous processes. 

Note that in our sample we find no evidence of AGNs forming in giant elliptical galaxies. Most possibly this is because such type of AGN would have only formed in the past. Today most giant ellipticals are in rich clusters of galaxies and are mostly inactive (or show evidence of activity under the form of some weak emission lines, but many more lines are missing). It is relatively easy, however, to extrapolate our results to what we could observe if such an AGN existed at higher reshifts. The SFH of this galaxy would be expected to show an even higher SFR in the past than in the LINERs, with a steeper positive $S_I$ and steeper negative $S_F$. Consequently, without a thorough analysis of its SFH, and being unable to establish its morphology, it would be very difficult to distinguish in this AGN between a quenching effect due to the SMBH and the high astration rate which is a normal feature for the formation of this type of galaxy. Taking this possibility in consideration, we believe that observations like those presented by \citet{Page12} are not necessarily incompatible with our results for the normal SFH of AGNs in NELGs of the nearby universe. 

\section*{Acknowledgments}

We like to acknowledge an anonymous referee for important comments that helped us produce a better version of our article. In particular, we owe the referee the definition for truncation of star formation. J.P. Torres-Papaqui also acknowledges PROMEP for support grant 103.5-10-4684, and DAIP-Ugto (0065/11). 
Funding for the SDSS and SDSS-II is provided by the Alfred P. Sloan Foundation. The SDSS is managed by the Astrophysical Research Consortium (ARC) for the Participating Institutions. The Participating Institutions are: the American Museum of Natural History, Astrophysical Institute Potsdam, University of Basel, University of Cambridge (Cambridge University), Case Western Reserve University, the University of Chicago, the Fermi National Accelerator Laboratory (Fermilab), the Institute for Advanced Study, the Japan Participation Group, the Johns Hopkins University, the Joint Institute for Nuclear Astrophysics, the Kavli Institute for Particle Astrophysics and Cosmology, the Korean Scientist Group, the Los Alamos National Laboratory, the Max-Planck-Institute for Astronomy (MPIA), the Max-Planck-Institute for Astrophysics (MPA), the New Mexico State University, the Ohio State University, the University of Pittsburgh, University of Portsmouth, Princeton University, the United States Naval Observatory, and the University of Washington.

\end{document}